\documentclass[%
 reprint,
 amsmath,amssymb,
 aps,
]{revtex4-2}

\usepackage{graphicx}
\usepackage{dcolumn}
\usepackage{bm}
\usepackage{hyperref}
\usepackage{cleveref}

\begin{document}

\preprint{APS/123-QED}

\title[Dielectric properties of the pore solution in cement-based materials]
  {Dielectric properties of the pore solution in cement-based materials}

\author{Tulio Honorio}
\affiliation{LMT, ENS Paris-Saclay, CNRS, Universit\'{e} Paris-Saclay, 94235 Cachan, France}%
 \email{tulio.honorio-de-faria@ens-paris-saclay.fr}

\author{Thierry Bore}%
\affiliation{School of Civil Engineering, The University of Queensland, Brisbane, Australia}

\author{Farid Benboudjema}
\affiliation{LMT, ENS Paris-Saclay, CNRS, Universit\'{e} Paris-Saclay, Cachan, France}

\author{Eric Vourc'h}
\author{Mehdi Ferhat}
\affiliation{ 
SATIE, UMR CNRS 8029, ENS Paris Saclay, Universit\'{e} Paris Saclay, Cachan, France}

\date{\today}

\begin{abstract}
Dielectric properties are intimately related to the content and composition of the liquid phase present in micro- and mesoporous materials.
The presence of ions is recognized to affect the dielectric response of electrolytes in a complex way that depends on the ion-water and ion-ion interactions and respective concentrations.
Pore solutions in cement-based materials are complex and age-dependent exhibiting different ion pair states, which make them an interesting candidate for the study of the complex pore solution found in geomaterials in general.
Here, we study the dielectric properties of bulk aqueous solutions representative of the pore solution found in cement-based materials using molecular dynamics simulations.
Broadband dielectric spectra and frequency-dependent conductivities are provided. 
A mean field upscaling strategy is deployed to compute the dielectric response at cement-paste scale.
In cement-based materials, the measurement of dielectric properties can be used for non-destructive monitoring of concrete structures conditions and to unravel details of the hierarchical pore structure of the material, notably at early-age when the microstructure of the material changes significantly. 
Our results can be used in the interpretation of high frequency electromagnetic methods, such as ground-penetrating radar, enhancing the quality of interpretation and improving the confidence in the corresponding results.
\end{abstract}


\maketitle


\section{Introduction}

High Frequency Electromagnetic (HF-EM) method have been deployed last decades to assess water content of soil and rocks \cite{topp_electromagnetic_1980,santamarina_soils_2001,bore_new_2018}, as well as man made geomaterials such as cementitious materials (e.g. \cite{klysz_determination_2007,rhim_electromagnetic_1998,laurens_influence_2002,van_beek_a._dielectric_2000,cataldo_tdr-based_2018}). 
Among these methods, one can quote L-band remote sensing \cite{marino_airborne_2001}, Ground Penetrating Radar (GPR) \cite{bungey_sub-surface_2004}, capacitive probe \cite{bore_capacitive_2013}, etc. 
The success of such methods is caused by the dipolar character of the water molecules resulting in a high permittivity in comparison to other phases such as solid particles or gas voids. 
However, evidence suggests that the frequency-dependent dielectric permittivity contains far more information than water content only: structure \cite{tang_review_2017}, density \cite{drnevich_time_2005}, mineralogy, cohesion forces on the pore solution \cite{escorihuela_influence_2007} and even material strength may leave distinct signatures that could potentially be quantified. 
Nevertheless, despite its great success, HF-EM methods are still confronted with theoretical challenges. Interaction of the pore solutions and solid phases leads to strong contributions to the EM properties. These interactions are not well understood at the moment: theoretical studies are urgently needed. 

As for other physical properties \cite{honorio_thermal_2018,bernard_multiscale_2003,stora_estimating_2008,wyrzykowski_numerical_2017}, micromechanics can be applied to upscale the dielectric permittivity of heterogeneous materials \cite{bruggeman_berechnung_1935,torquato_random_2002}.
However, important information concerning the behavior of the constituent phases is required to provide reliable predictions of the effective dielectric permittivity of nanoporous materials, is lacking.
Dielectric properties are recognized to be dependent on pore size \cite{ballenegger_dielectric_2005,bonthuis_dielectric_2011} and pore solution composition \cite{rinne_ion-specific_2014,rinne_dissecting_2014,schaaf_dielectric_2015}.
The relative static dielectric permittivity $\epsilon(0)$ of aqueous solutions are in general one order of magnitude larger (e.g. for bulk water is \cite{malmberg_dielectric_1956} $\epsilon(0)$=80.103 at 20$^{\circ}$C ) than the dielectric permittivity of solid phases. 
Therefore, the precise quantification of the dielectric response of the pore solution is crucial to upscale the dielectric properties and to interpret the experimental spectra of porous materials.

In cement-based materials, the composition of the pore solution is crucial to the stability and assemblage of phases in these materials \cite{vollpracht_pore_2016}. Therefore, cement hydration processes and chemical alterations of cement-based materials (due, for instance, to durability issues) are closely related to the pore solution composition.
The composition of the pore solution depends on the composition of the binder (e.g. ordinary Portland cement, alternative cements, and Supplementary Cementitious Materials (SCM) a), age and temperature \cite{lothenbach_thermodynamic_2006,lothenbach_thermodynamic_2008,vollpracht_pore_2016}.
Atomistic simulations have been successfully used to unveil the details of water interaction in cement-based materials, specially in calcium silicate hydrates (e.g.\cite{pellenq_engineering_2008,youssef_glassy_2011,bonnaud_interaction_2016,ozcelik_nanoscale_2016,masoumi_effective_2019,honorio_monte_2019}).
Recently, the authors have studied for the first time the dynamics and structure of bulk aqueous solutions representative of the pore solution found in cement-based materials through molecular dynamics simulations \cite{honorio_pore_2019}.
The repercussions of the dynamics and structure differences observed in these pore solution on the dielectric response are still to be quantified.

Here, the dielectric properties of the pore solution of cement-based materials are computed taking into account the variability of ionic concentrations in aqueous solutions.
To do so, we perform MD simulations to compute the dielectric response of the pore solutions.
The composition of the pore solutions, atomic configurations and force fields, which are recalled in this article, are taken from Honorio et al. \cite{honorio_pore_2019}.
For the first time, the dielectric spectra of cement-based materials pore solutions are computed accounting for the age-dependent composition variations.
We observe that the Cole-Cole equation fits well the dielectric spectra obtained for all pore solutions.
Finally, we quantify the effect of the variability of the dielectric response of the pore solution on the effective response of cement-based materials using Monte Carlo Micromechanics \cite{honorio_multiscale_2019}.
Our results contribute to a better understanding of the electromagnetic behavior of cement-based materials and can be readily used in the interpretation of dielectric probing of cement-based materials, enhancing the performance of HF-EM testing and improving the confidence in the corresponding results. 

\section{Theory and Calculation}

\subsection{Dielectric response and conductivity from molecular simulations }

For a system with $n$ particles $i$ with charge $q_i$, the total system polarization $\vec{P}$ is defined as the sum of the the polarization (or dipole moment) $\vec{\mu_i}(t)$ of the each particle $i$ at time $t$ \cite{praprotnik_molecular_2005}:

\begin{equation}
\vec{P}(t)=\sum_{i=1}^n \vec{\mu_i}(t)=\sum_{i=1}^n q_i \vec{r_i}(t) \label{eq:tot_P}
\end{equation}

\noindent where $\vec{r_i}(t)$ is the position of the (center of) particle $i$.

The total system polarization $\vec{P}$ is related to the electric field $\vec{E}$ via the complex frequency-dependent dielectric susceptibility $\chi(f)=\chi'(f)-i \chi''(f)$ by

\begin{equation}
\vec{P}(f)=\chi(f)\epsilon_0 \vec{E}(f)  \label{eq:lin_P_eps}
\end{equation}

\noindent  where $\epsilon_0$ is the vacuum permittivity.
The dielectric susceptibility $\chi(f)$ can be computed from the auto-correlation of equilibrium polarization fluctuations using \cite{kohler_liquid_1972,rinne_dissecting_2014}:

\begin{equation}
\chi(f)=-\frac{1}{3V k_B T \epsilon_0}\int^{\infty}_0 e^{-2 \pi i f t} \left\langle \vec{P}(0). \dot{\vec{P}}(t) \right\rangle dt \label{eq:autocorr_suscep}
\end{equation}

\noindent where $V$ and $T$ are the volume and temperature of the system, respectively;  $k_B$ is the Boltzmann constant and $\dot{\vec{P}}(t)$ is the time-derivative of the total polarization.

For a salt aqueous solution, the total polarization is the sum of the contributions of the water $\vec{P}_W$ and ionic $\vec{P}_I$ polarizations: $\vec{P}=\vec{P}_W+\vec{P}_I$.

The ionic current $\vec{J}_I$ is related to the ionic polarization by: $\vec{J}_I(t) =\dot{\vec{P}}_I(t)$.
This ionic current $\vec{J}_I$ is linked to the electric field $\vec{E}$ via the frequency-dependent ionic conductivity $\sigma(f)=\sigma'(f)-i \sigma''(f)$ by

\begin{equation}
\vec{J}_I(f)=\sigma(f)\epsilon_0 \vec{E}(f)  \label{eq:lin_J_eps}
\end{equation}

\noindent The frequency-dependent ionic conductivity can be computed using the cross correlations \cite{caillol_theoretical_1986,rinne_dissecting_2014}: 

\begin{equation}
\sigma(f)=\frac{1}{3V k_B T }\int^{\infty}_0 e^{-2 \pi i f t} \left\langle \vec{J}_I(0). \dot{\vec{P}}(t) \right\rangle dt \label{eq:autocorr_cond}
\end{equation}

Following Rinne et al. \cite{rinne_dissecting_2014,rinne_ion-specific_2014}, we define the following auto- and cross-correlations functions of the water polarization and ionic current:

\begin{equation}
\phi_W(t)=\frac{\left\langle \vec{P}_W(0). \vec{P}_W(t) \right\rangle}{3V k_B T \epsilon_0} \label{eq:phi_W}
\end{equation}

\begin{equation}
\phi_{IW}(t)=\frac{\left\langle \vec{P}_W(0). \vec{J}_I(t) -\vec{J}_I(0). \vec{P}_W(t) \right\rangle}{3V k_B T \epsilon_0} \label{eq:phi_IW}
\end{equation}

\begin{equation}
\phi_I(t)=\frac{\left\langle \vec{J}_I(0). \vec{J}_I(t) \right\rangle}{3V k_B T \epsilon_0} \label{eq:phi_I}
\end{equation}

\noindent from which the water, ion and ion-water interaction contributions on dielectric susceptibility and frequency-dependent ionic conductivity can be quantified.

Accordingly, the regularized susceptibility $\Delta \chi(f)$ can be decomposed into three separate contributions: 

 \begin{equation}
\Delta \chi(f)=\chi_{W}(f)+\chi_{IW}(f)+\Delta \chi_I(f) \label{eq:delta_susc}
\end{equation}

\noindent where 

\begin{equation}
\chi_{W}(f)=\phi_{W}(0)-i 2 \pi f \int^{\infty}_0 e^{-2 \pi i f t} \phi_{W}(t) dt \label{eq:chi_W}
\end{equation}

\begin{equation}
\chi_{IW}(f)=- 2 \int^{\infty}_0 e^{-2 \pi i f t} \phi_{IW}(t) dt \label{eq:chi_IW}
\end{equation}

\begin{equation}
\Delta \chi_{I}(f)=-\frac{i}{2 \pi f}\int^{\infty}_0 \left( e^{-2 \pi i f t} -1 \right) \phi_{I}(t) dt \label{eq:chi_I}
\end{equation}

\noindent The frequency-dependent ionic conductivity can be decomposed into the two terms:

\begin{equation}
\sigma(f)=\sigma_{IW}(f)+\sigma_I(f) \label{eq:sigm_sum}
\end{equation}

\noindent with

\begin{equation}
\sigma_{IW}(f)=-i2 \pi f \epsilon_0 \int^{\infty}_0 e^{-2 \pi i f t} \phi_{IW}(t) dt \label{eq:sig_IW}
\end{equation}

\begin{equation}
\sigma_{I}(f)= \epsilon_0 \int^{\infty}_0 e^{-2 \pi i f t} \phi_{I}(t) dt \label{eq:sig_I}
\end{equation}

\noindent Therefore, the static conductivity can be computed via $\sigma(f=0)=\sigma_I(f=0)= \epsilon_0 \int^{\infty}_0  \phi_{I}(t) dt$.

\subsection{Molecular modeling of the pore solutions}

\subsubsection{Pore solutions in cement-based materials: variability and definition of scenarios}

Here, we study the bulk aqueous solutions mimicking the pore solutions in cement-based materials as in ref. \cite{honorio_pore_2019}.
One can legitimately raise the question of the relevance of analyzing bulk solutions to get insights on pore solution behavior.
As previously discussed \cite{honorio_pore_2019}, the most significant part of the dynamics of water and ions in pore solution within cement-based materials can be described by the dynamics of the fluid portions less affected by pore walls.
The ionic transport occurring in the pore zones affected by the interface (Stern layer) is reported to be negligible when compared to transport in the bulk pore, especially for aqueous solutions with a concentration similar to pore solutions in cement-based materials \cite{samson_modeling_2005,revil_ionic_1999}.
Furthermore, atomistic simulations of water confined in mesopores show that for a large portion of these pores the fluid behaves as a bulk fluid (e.g \cite{honorio_anomalous_2019}).
The smaller pore size found in cement-based materials is the inter-layer calcium silicate hydrates (C-S-H) pores of approximately 0.95~nm \cite{muller_use_2013}, but C-S-H gels exhibit also a mesoporosity ranging from few up to tens of nanometers.
Specific techniques \cite{ballenegger_dielectric_2005,bonthuis_dielectric_2011} are required to study the water confined in pore as small as C-S-H inter-layer pores.
Regarding larger pores, though, molecular simulations of C-S-H and other mesoporous materials show that for pores as small as 5 nm, in a major portion of the pore the fluid shows bulk-like behavior \cite{honorio_monte_2019,honorio_anomalous_2019,honorio_molecular_2019}.
These arguments show that a large portion of the fluids in cement-based materials porosity is expected to exhibit bulk-like behavior.
Furthermore, most HF-EM remote sensing method is carried out in a frequency range around 1~GHz.
In this frequency range, the effect of free water (or bulk water) is predominant. It was shown by previous studies that adsorbed water (chemically or hydrogen-bond) will manifest at for lower frequency \cite{miura_microwave_1998,hager_monitoring_2004}. 
Therefore, critical physical insights that can be relevant to the understanding of the dielectric response of cement-based materials might be inferred from the analysis of bulk solutions representative of the pore solutions.

\begin{figure*}
 \includegraphics[width=15cm]{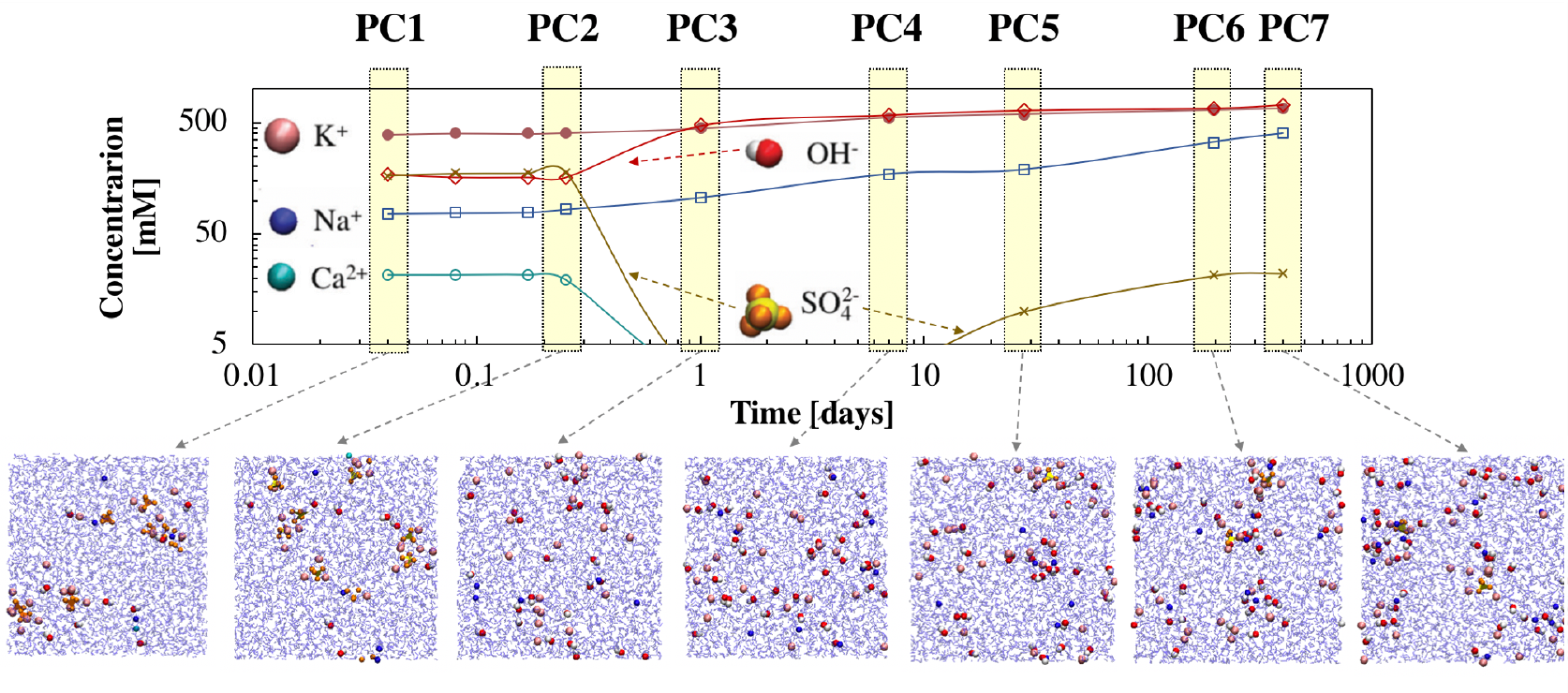}
 \caption{Evolution of the concentrations of the main ionic species (concentrations exceeding 5~mM) in ordinary Portland cement (PC) pore solutions. The scenarios PC1-PC7 are retained (snapshots of atomic configurations at the bottom) to study the age-dependency of the pore solution composition.}
  \label{fgr:F1_comp_var}
\end{figure*}

The composition of the pore solutions is based on the experimental data provided by Lothenbach et al. \cite{lothenbach_influence_2008} for an ordinary Portland cement (PC) systems in a formulation with a water-to-cement $w/c$ ratio of 0.4.
Seven scenarios PC1-PC7 corresponding to various ages are chosen to be simulated. The species with a concentration below 5 mM are not taken into account.
Table \ref{tab:Conc_Pore} shows the compositions of the pore solutions studied here and the associated age in ordinary Portland cement (PC) systems. 
The ionic concentrations were adjusted to comply with the electroneutrality of the simulated system.
Figure \ref{fgr:F1_comp_var} shows the evolution of the concentration of the ionic species plotted as a function of the age.
For each age, we present a snapshot of the atomic configurations in a system equilibrated in a canonical simulation at 300K.

\begin{table}
\caption{\label{tab:Conc_Pore} Composition of the simulated solutions (as in Honorio et al. \cite{honorio_pore_2019} based on the experimental data provided by Lothenbach et la. \cite{lothenbach_influence_2008}). Adjustments in the ionic concentration were made to ensure the electroneutrality of the system. }
\begin{ruledtabular}
\begin{tabular}{c|c|cccccc}
  Sim.    & $t$        & H$_2$O     & Na$^+$     & K$^+$     & Ca$^{2+}$     & SO$_4^{2-}$     & OH$^{-}$      \\ 
             &  [days]    &                     &                     &       &                      &                      &                    \\ 
\hline
PC1  &  0.04     &     2767     &     4     &     20    &     1     &     8         &     10        \\ 
PC2  &  0.25     &     2767     &     4     &     20     &     1     &     9         &     8      \\ 
PC3  &  1         & 2767         &     5     &     20     &     0    &     0         &     25      \\ 
PC4 &  7         & 2767         &     9     &     24    &     0    &     0         &     33      \\ 
PC5 &  28      & 2767         &  9     &     27        &     0    &     1         &     34     \\ 
PC6  &  197      & 2767         & 12     &     29    &     0    &     2         &     37     \\ 
PC7  &  400      & 2767        & 16 &     30     &     0    &     2         &     42          \\ 
\end{tabular}
\end{ruledtabular}
\end{table}

\subsubsection{Force fields and simulation details}

A detailed description of the force field parameters is provided in a previous work \cite{honorio_pore_2019}.
The interactions among all the species in the system are modeled by a sum of non-bonded (van der Waals and electrostatic) and bonded (angle and bonds for OH groups) interactions.  
The van der Waals interactions are described either by the Lennard-Jones (12-6) potential for interactions between water (SPC/E \cite{berendsen_missing_1987}), hydroxide and monatomic ions or by the Buckingham potential for interactions involving sulfates. The force field parameters are provided as Supporting Information.
Long-range van der Waals interactions are treated with tail corrections.
Coulomb potential describes the electrostatic contribution. Ewald sum method is used to cope with long-range electrostatic interactions.
Water and hydroxide bonds and water angles are constrained by SHAKE algorithm.
Sulfate ions are constrained using RIGID algorithm. 

We use LAMMPS \cite{plimpton_fast_1995} to perform the simulations.
Initial atomic configurations were obtained by randomly placing the ions in the simulation box following the composition in Tab.\ref{tab:Conc_Pore}. 
Next, a microcanonical (NVE) simulation with an imposed maximum displacement of 0.01~\r{A} per atom was performed during 1 ps to cope with possible overlapping atoms.
Then, the system was equilibrated during 0.25~ns at 300 K in a canonical simulation (NVT).
Nos\'{e}-Hoover thermostat is used for equilibration and production with a damping parameter of 100 timesteps.
The dipole moments of water and all ions were computed separately each 25~fs in an NVT simulation with a 1~fs timestep during 1.0~ns. 
As shown in the Supporting Information, 1 ns seems enough to sample the fluctuations of the water polarization and ionic currents.
The effects of finite size and fictitious forces from Nos\'{e}-Hoover thermostat are discussed in the Supporting Information.

\section{Results and Discussion}

\begin{figure*}
 \includegraphics[width=12cm]{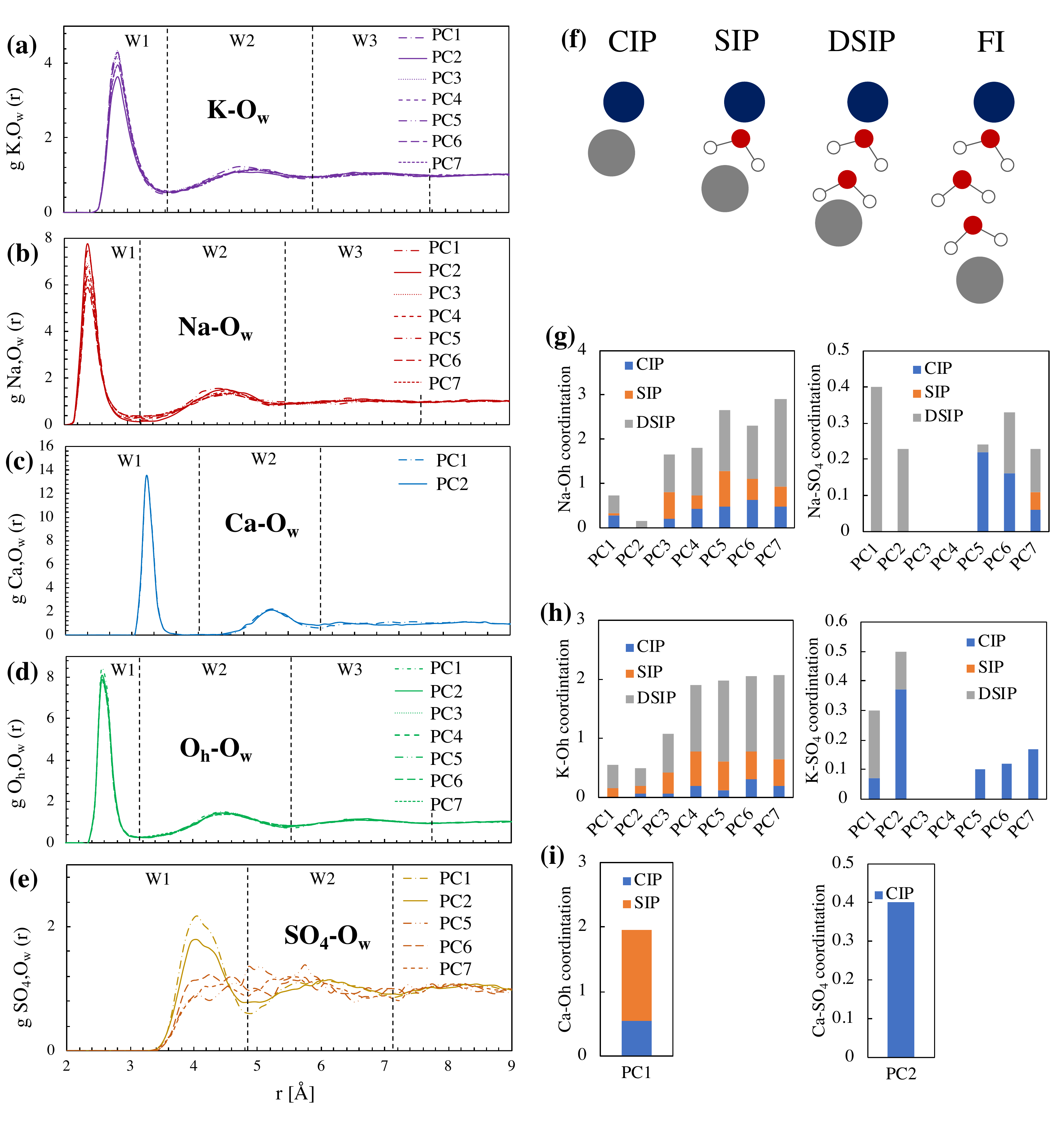}
 \caption{Ion hydration shells: radial distribution functions of (a) K$^+$, (b) Na$^+$, (c) Ca$^{2+}$, (d) OH$^{-}$ and (e) SO$_4^{2-}$ -O$_w$ pairs. W1, W2 and W3 denotes 1, 2 and 3 hydration shells, respectively.
 Ion pair states: (f) schematic representation of contact ion pairs (CIP), single solvent-separated ion pairs (SIP), doubly solvent-separated ion pairs (DSIP), and free ions (FI) \cite{rinne_dissecting_2014}. Anions coordinated by cations: OH$^{-}$ and SO$_4^{2-}$ coordinated by (g) Na$^+$,  (h) K$^+$, and (i) Ca$^{2+}$. The ion-ion radial distribution functions used to compute these histograms are detailed in the Supporting Information. }
  \label{fgr:F2_RDF_Count_Ion}
\end{figure*}

Figure \ref{fgr:F2_RDF_Count_Ion} (a)-(e) shows the water solvation shells of the ions present in the pore solutions PC1-PC7.
As discussed previously \cite{honorio_pore_2019}, cation-water and hydroxide-water structuration are similar in all pore solutions studied; whereas sulfates-water pair exhibits structure variations according to the pore solution. 
We compute the ion pair states following contact ion pairs (CIP), single solvent-separated ion pairs (SIP) and doubly solvent-separated ion pairs (DSIP) configurations depicted in Fig. \ref{fgr:F2_RDF_Count_Ion} (f).
DSIP configurations are prevalent in most of the cation-hydroxide pairs while CIP configurations are prevalent in most of the cation-sulfate pairs.

Three main stages can be identified regarding the ionic composition of the pore solution for a given cement system  \cite{honorio_pore_2019}: 
\begin{itemize}
\item \textit{very early age} up to setting (first hours): PC1 and PC2; 
\item \textit{early-age} related to property development (within the first days): PC3 and PC4;
\item \textit{late stage} associated and the service life of cement-based materials: PC5 to PC7.
\end{itemize}
\noindent such classification has been useful in linking the dynamics of ions to the electrical properties of cement-based materials \cite{honorio_multiscale_2019}.
As shown in Fig. \ref{fgr:F2_RDF_Count_Ion} (g) and (h), ion pair states seems to obey this classification in most cases.

\begin{figure*}
 \includegraphics[width=16cm]{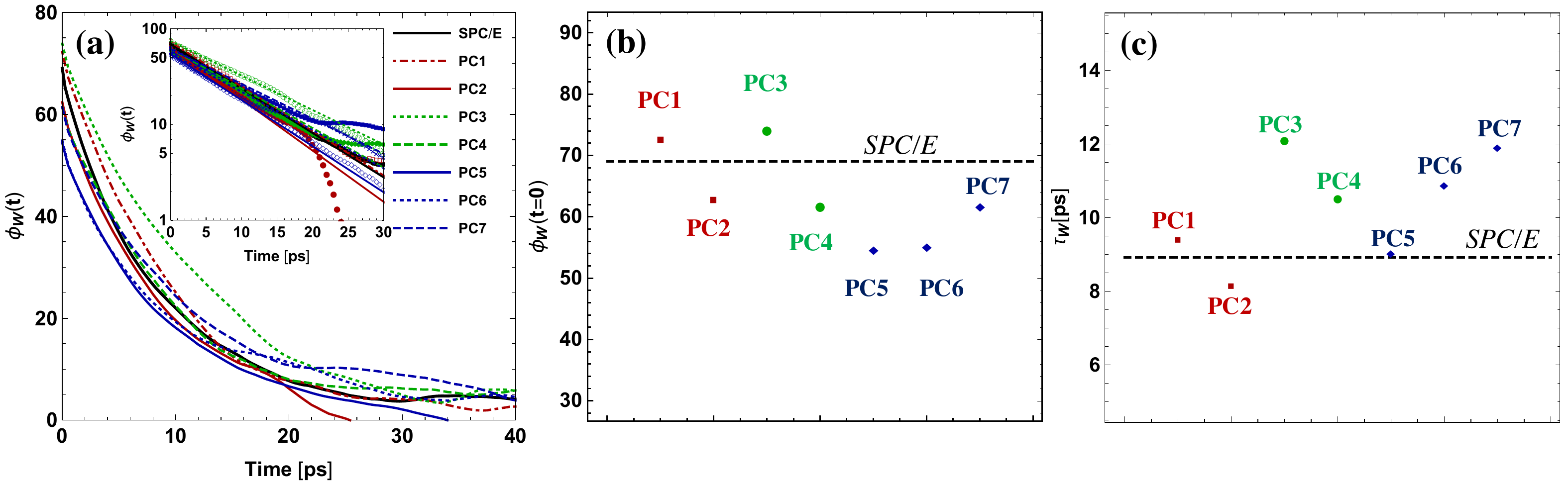}
 \caption{(a) Auto-correlation function of the water polarization $\phi_W(t)$ of the pore solutions PC1-PC7 and SPC/E water. The inset shows Debye fits in a Log-plot.
 (b) $\phi_W(t=0)$ for pore solutions PC1-PC7 and SPC/E water.
 (c) Characteristic relaxation time obtained from Debye fits.}
  \label{fgr:F3_Corre_W}
\end{figure*}

Figure  \ref{fgr:F3_Corre_W} (a) shows the auto-correlation function $\phi_W(t)$ of the water polarization for all the pore solutions studied and for (pure) SPC/E water.
This auto-correlation function is the most relevant for the computation of the dielectric spectra \cite{rinne_dissecting_2014}.
In all cases, exponential decay is observed.
Due to noise, $\phi_W(t)$ exhibit negative values for times exceeding 25 to 60 ps for all solutions.
We fit $\phi_W(t)$ with the expression $\phi_W(t=0) = Exp[-t/\tau_D]$, where $\tau_D$ is the characteristic relaxation time. The values of $\phi_W(t=0)$ and  $\tau_D$ are plotted for each pore solution and for SPC/E water in Figs.  \ref{fgr:F3_Corre_W} (b) and (c), respectively.
No clear tendency is observed in the values for the pore solution concerning the three stages classification or the similarity with respect to SPC/E water values.

\begin{figure*}
\center
 \includegraphics[width=15cm]{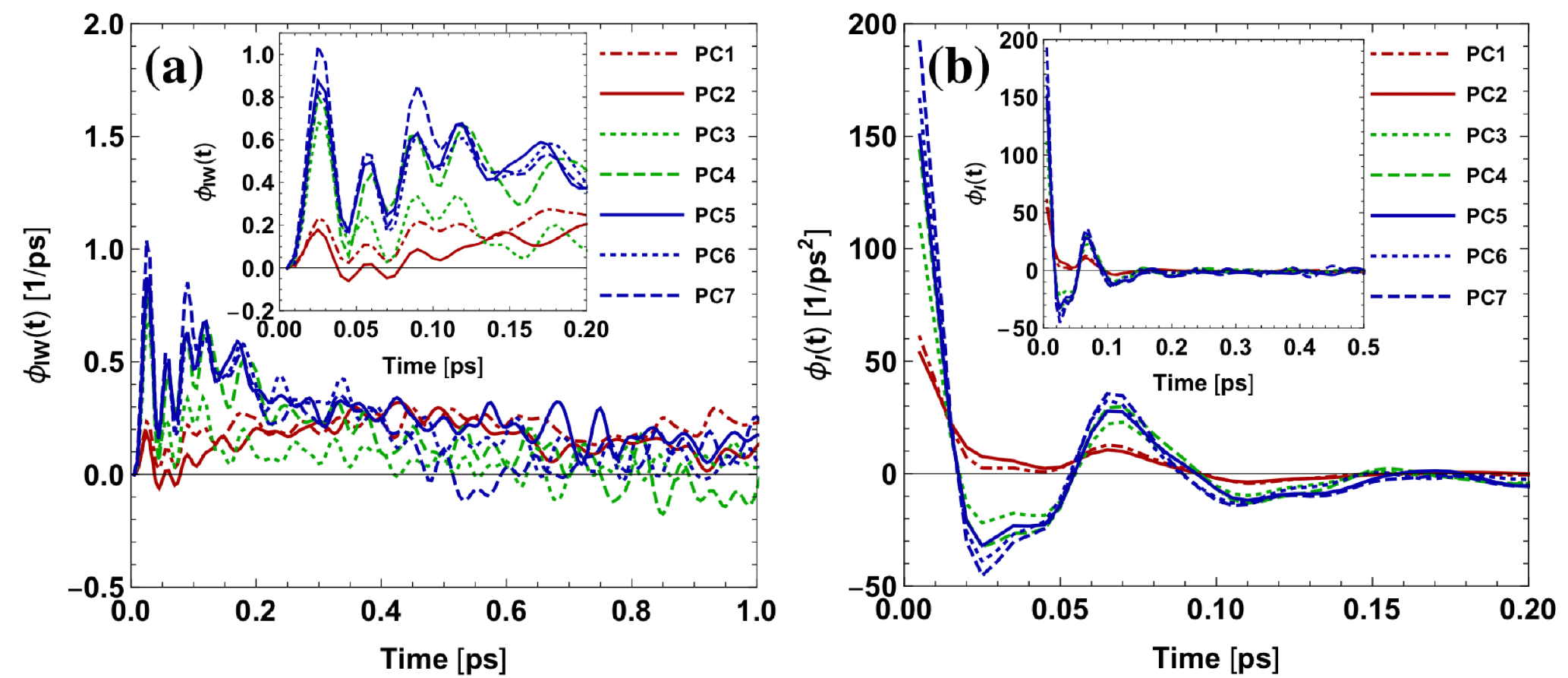}
 \caption{(a) Cross-correlation function $\phi_{IW}(t)$ of water polarization and ionic current for pore solutions PC1-PC7 and SPC/E water. The inset zooms in smaller timescales. (b) Auto-correlation function  of the ionic current ($\phi_I(t)$) for pore solutions PC1-PC7 and SPC/E water. The inset shows the the decay at longer timescales.}
  \label{fgr:F4_phiIW_phiI}
\end{figure*}

Ion-water cross-correlation $\phi_{IW}(t)$ and ion auto-correlation $\phi_I(t)$ functions are presented in Figs. \ref{fgr:F4_phiIW_phiI} (a) and (b), respectively. 
Both functions present a high-frequency oscillatory behavior with the oscillations being more strongly damped in the case of $\phi_I(t)$.
Strong dampening has been also observed in NaCl, NaI, NaBr and NaF aqueous solutions \cite{rinne_dissecting_2014}.
The ion auto-correlation functions $\phi_I(t)$ of the pore solution at the very early-age (PC1 and PC2) show an oscillatory behavior with lower amplitude when compared to the other solutions; the larger amplitudes are observed with the solution at later stages (PC5-PC7).
As observed to other aqueous solutions (NaCl, NaI, NaBr and NaF),  ion-water cross-correlation function $\phi_{IW}(t)$ present an initial steep increase followed by an initial decay and a subsequent slower relaxation.
Compared to the other simpler aqueous solutions, the pore solution shows persistent oscillations even for times above 1 ps.

\subsection{Dielectric spectra}

\begin{figure*}
 \includegraphics[width=15cm]{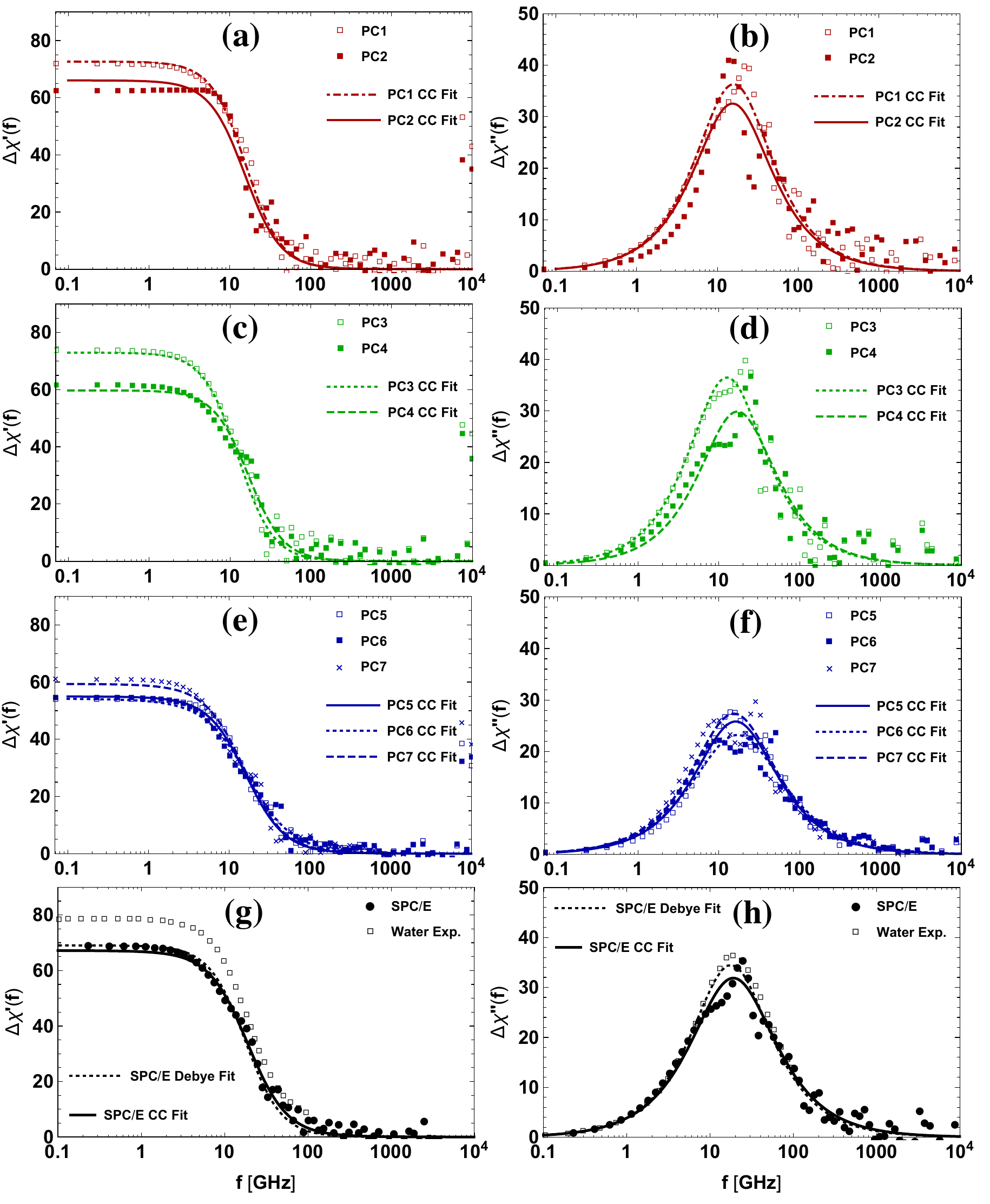}
 \caption{Dielectric spectra of the pore solutions: real (left) and imaginary (right) parts of the dielectric susceptibility:  $\Delta \chi(f)'$ and $\Delta \chi(f)''$, respectively.  (a)-(b) very early-age, (c)-(d) early-age, and (e)-(f) late ages pore solutions. Full lines are Cole-Cole fits of the MD results. 
 For comparison, the experimental spectrum obtained by Buchner et al.  \cite{buchner_dielectric_1999} is also shown (g)-(h).}
  \label{fgr:F5_suscep_Tot_CC}
\end{figure*}

\begin{figure*}
 \includegraphics[width=15cm]{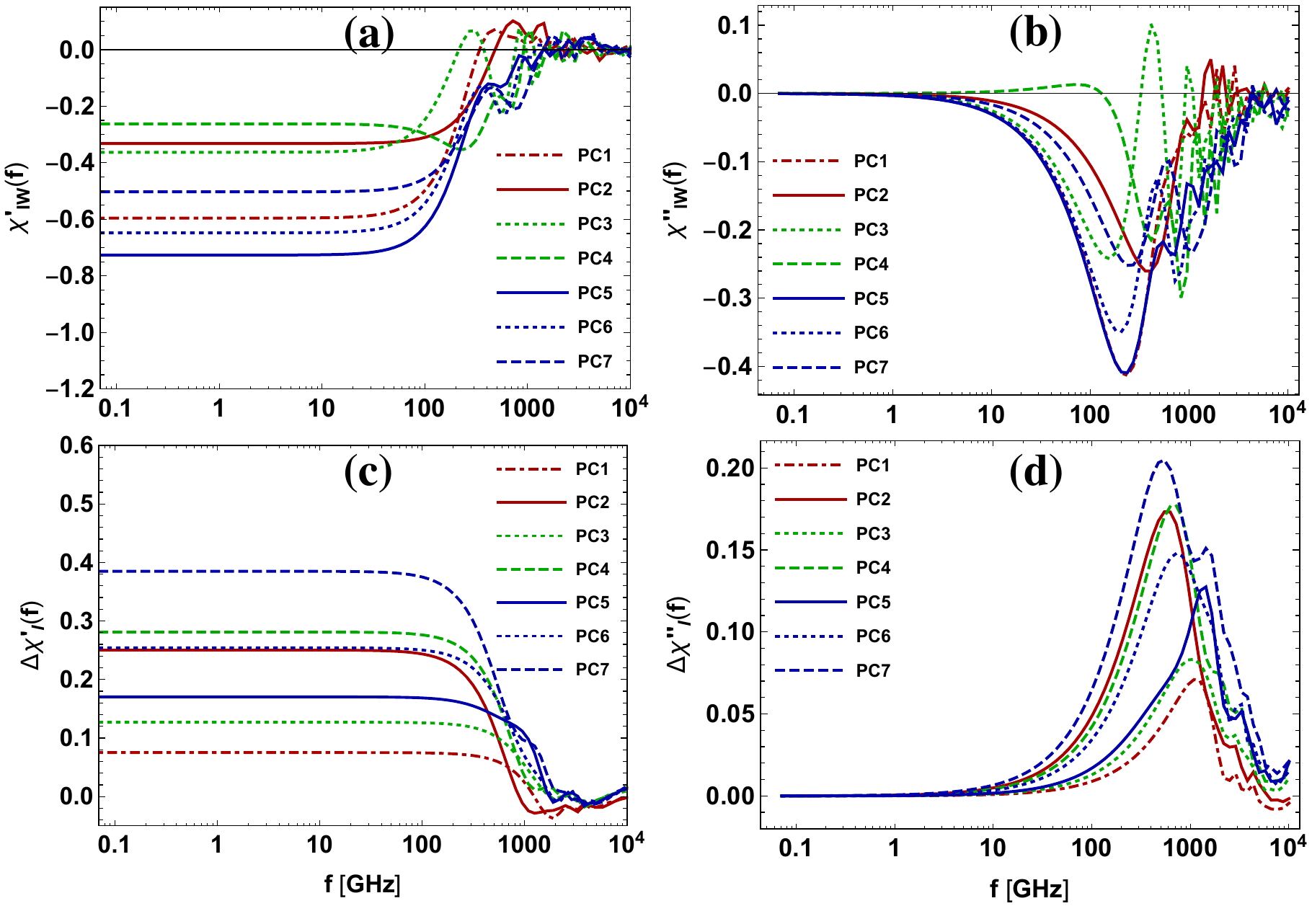}
 \caption{Dielectric spectra of the pore solutions PC1-PC7: (a) real $\chi_{IW}(f)'$ and (b) imaginary $\chi_{IW}(f)''$ parts of the ion-water dielectric contribution; 
 (c) real $\Delta \chi_{I}(f)'$ and (d) imaginary $\Delta \chi_{I}(f)''$ parts of the DC-conductivity dielectric ion contribution.}
  \label{fgr:F6_suscep_IW_I}
\end{figure*}

The dielectric spectra of the pore solutions and SPC/E including all contributions are shown in Fig.
\ref{fgr:F5_suscep_Tot_CC} together with the respective Cole-Cole fits.
The water contribution is predominant (see the Supporting Information for the results without ion contributions). 
The ion-water dielectric contributions $\chi_{IW}(f)$ on the dielectric spectra of the pore solutions are shown in Fig. \ref{fgr:F6_suscep_IW_I} (a) and (b), respectively. 
The real and the imaginary parts of the DC-conductivity dielectric ion contribution $\Delta \chi_{I}(f)$ for all pore solution are shown in  Fig. \ref{fgr:F6_suscep_IW_I} (c) and (d), respectively.

To fit the normalized dielectric spectrum we use Cole-Cole model \cite{cole_dispersion_1941,rinne_dissecting_2014}:

\begin{equation}
\Delta \chi(f)=\frac{\epsilon_{CC}-\epsilon_{\infty}}{1+\left(i 2\pi f \tau_{CC} \right)^{1-\alpha}}+\epsilon_{\infty}-1  \label{eq:Cole_Cole}
\end{equation}

\noindent where $\epsilon_{CC}$ and $\tau_{CC}$ are, respectively, the Cole-Cole amplitude and characteristic relaxation time; and, $\alpha$ is an exponent parameter ranging from 0 to 1 ($\alpha$=0 corresponding to Debye model). 
We adopt $\epsilon_{\infty}$=1 since the force fields employed in MD simulation do not comprise atomic polarization.
To fit simultaneously the real and imaginary parts of the dielectric spectrum, we minimize the function $\Delta \chi_i(f) \bar{\Delta \chi_i(f) }$, where $\Delta \chi_i(f) $ refers to the spectrum data to be fitted and $\bar{.}$ denotes the complex conjugate.

The parameters of Cole-Cole fits are gathered in Tab. \ref{tab:CC_fits}. 
Water simulation and experimental data from the literature are also provided.
The static dielectric permittivity of SPC/E water is known to be inferior to the experimental value.  
We obtain 69.0 for SPC/E water, which is in agreement with other values from the literature (e.g.\cite{smith_computer_1994}  67~$\pm$~10).
In spite of its simplicity, the SPC/E model shows a remarkable agreement with experiments.

\begin{figure*}
 \includegraphics[width=12cm]{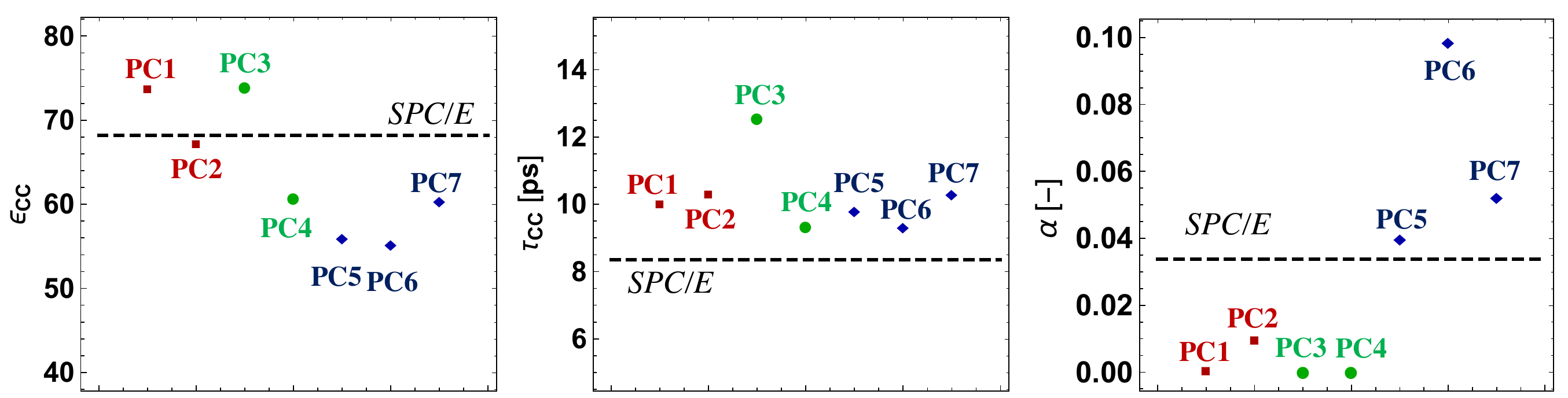}
 \caption{Parameters of Cole-Cole fits from Fig. \ref{fgr:F5_suscep_Tot_CC}.}
  \label{fgr:F7_CC_par}
\end{figure*}

Figure \ref{fgr:F7_CC_par} shows the parameters of Cole-Cole fits.
The amplitude of Cole-Cole fits is inferior to that of SPC/E water for pore solution at late ages (PC5-PC7).
Cole-Cole relaxation times of all pore solutions exceed that of SPC/E water, with very early-age (PC1-PC2) and late ages (PC5-PC7) pore solution exhibiting similar $\tau_{CC}$.
The $\alpha$ parameter of very early-age (PC1-PC2) and early-age (PC3-PC4) pore solutions are closer to 0 (i.e. a Debye-like behavior) than the values of late ages solutions (PC5-PC7). 

In agreement with our results, recent experimental measurements of the dielectric response of a concrete pore solution show that the real part remains almost constant for frequencies up, at least, to 2~GHz \cite{guihard_permittivity_2019}.
The same authors also showed that a Debye model can describe fairly well the dielectric spectra of the pore solution, which is also consistent with our results that show values of $\alpha$ close to zero.

\begin{table}
\small
\caption{\label{tab:CC_fits} Cole-Cole fit parameters from MD simulations of pore solution and SPC/E water, and data from the literature on water.}
\begin{ruledtabular}
\begin{tabular}{c|ccc}
 &     $\epsilon_{CC}$ &      $\tau_{CC}$  &     $\alpha$  \\ 
\hline
PC1   &     73.6  &   9.97  &  0.000  \\
PC2   &     67.1  &   10.26  &  0.009  \\
PC3   &     73.9  &  12.55  &  0.000  \\
PC4   &     60.7  &   9.33  &  0.000  \\
PC5   &     56.0  &   9.80  &  0.040  \\
PC6   &     55.2  &   9.33  &  0.098  \\
PC7   &     60.4  &   10.30  &  0.052  \\
 \hline 
SPC/E \footnotemark[1]    &     68.2  &   8.35  &  0.033  \\
SPC/E \footnotemark[2]   &     78.4  &   8.27  &  0.000  \\
SPC/E \footnotemark[3]   &     69.9  &   10.72  &  0.014  \\
\end{tabular}
\end{ruledtabular}
\footnotetext[1]{This work.}
\footnotetext[2]{Buchner et al. \cite{buchner_dielectric_1999} experiments.}
\footnotetext[3]{Rinne et al. \cite{rinne_dissecting_2014} MD simulations.}
\end{table}

\subsection{Electrical conductivity}

\begin{figure*}
 \includegraphics[width=15cm]{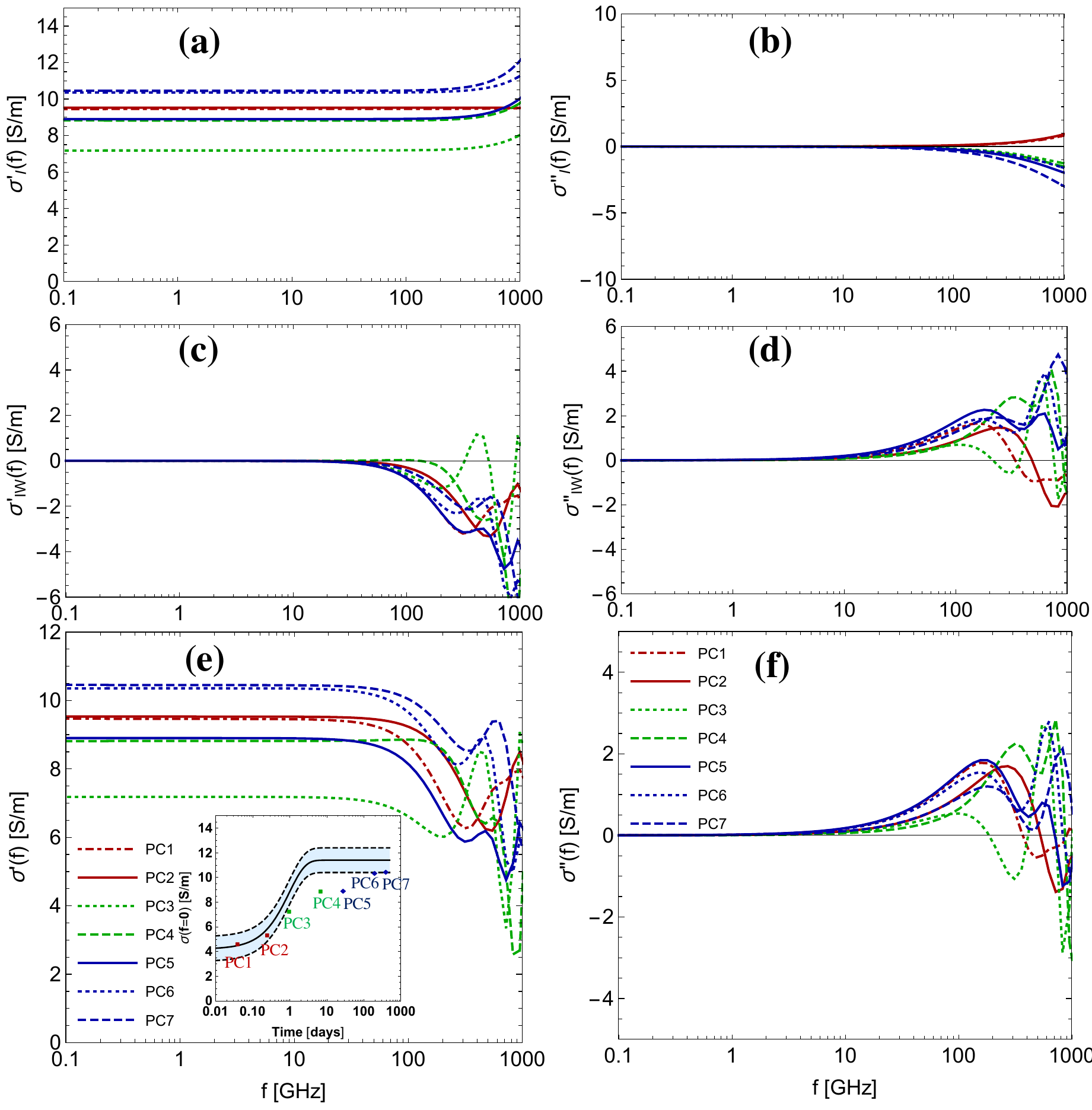}
 \caption{Frequency-dependent conductivity of the pore solutions: real (left) and imaginary (right) parts of the dielectric susceptibility: (a)-(b) ion contribution, (c)-(d) ion-water contribution, and (e)-(f) the total electrical conductivity of the pore solutions.
 The inset in (e) shows the static part of the electrical conductivity compared with the master curve weighted with respect to the w/c of 0.4 (full line) and the corresponding variability of approximately 1 S/m (bounds defined by the dashed curves) as discussed by Honorio et al. \cite{honorio_multiscale_2019}. }
  \label{fgr:F8_sigma}
\end{figure*}

Figure \ref{fgr:F8_sigma} shows the real and imaginary parts of the frequency-dependent conductivity of the pore solutions.
The ion-ion and ion-water contributions are shown.
Our results are in agreement with the static conductivity reported for ordinary PC systems with a water-to-cement ratio of $w/c=$0.4, which can be described by a single master curved weighted by the $w/c$ as discussed by Honorio et al. \cite{honorio_multiscale_2019}. 
The conductivities of the pore solutions are similar to the static value up to frequencies on the order of 10 GHz. AT higher frequencies both ion-ion and ion-water contributions exhibit oscillating responses with respect to the frequency.

\subsection{Upscaling the effects of the variability pore solution dielectric response}

The homogenization of the dielectric permittivity and electrical conductivity is analogous to the homogenization of the thermal conductivity and diffusivity \cite{torquato_random_2002}.
Bruggeman (or the self-consistent) scheme has been shown to provide a good estimate of the effective electrical conductivity of cement-pastes by capturing the transition from a liquid to a solid matrix during cement hydration \cite{honorio_multiscale_2019}.

In a polycrystalline-like morphology, for a $N$-phase heterogeneous materials with $N$ isotropic equiaxed inclusions randomly distributed in representative elementary volume, the Bruggeman (or Self-Consistent) estimate of the effective electrical permittivity $\epsilon^{SC}$ can be computed from the implicit formula \cite{torquato_random_2002}:

\begin{equation}
\sum_{r=1}^N f_r \frac{\epsilon_{r}-\epsilon^{SC}}{\epsilon_{r}+2\epsilon^{SC}}=0 \label{eq:SC_iso}
\end{equation}

\noindent for all phases $r$ with volume fraction $f_r$ and dielectric permittivity $\epsilon_{r}$.

To upscale the effective permittivity $\epsilon = \Delta \chi+1$ of the cement paste we use the Cole-Cole fits with the parameters in Tab. \ref{tab:CC_fits} to describe the dielectric permittivity of each pore solution $\epsilon_{PS}$.
Powers cement hydration model \cite{powers_studies_1946} is used to estimate the evolution of the capillary porosity (i.e. the volume fraction associated to the pore solution phase) in the system:

\begin{equation}
f_{PS}(\xi , w/c)=\phi^0_{cap}(w/c)-1.32(1-\phi^0_{cap}(w/c)) \xi \label{eq:phi_Powers}
\end{equation}

\noindent where the degree of hydration $\xi$ can be described by the sigmoid $\xi=0.9 \left[1 - \text{Exp} \left(-t/7 \right)\right]$ for $t$ in days as in ref. \cite{honorio_multiscale_2019}, and $\phi^0_{cap}(w/c)=(w/c)/\left(\rho_w / \rho_c +(w/c) \right)$ is the initial porosity computed from the $w/c$ and densities of water and cement ($\rho_c$ and $\rho_w$, respectively).
It is worth to note that this approach does not take into account interface processes that will dominate the spectra around kHz and MHz range \cite{miura_microwave_1998,hager_monitoring_2004}. 
Nevertheless, as far as the authors know, it is the first attempt to use effective medium theory with realistic data on pore solution dielectric response for the modeling of HF-EM properties of cement paste. 

\begin{figure*}
 \includegraphics[width=15cm]{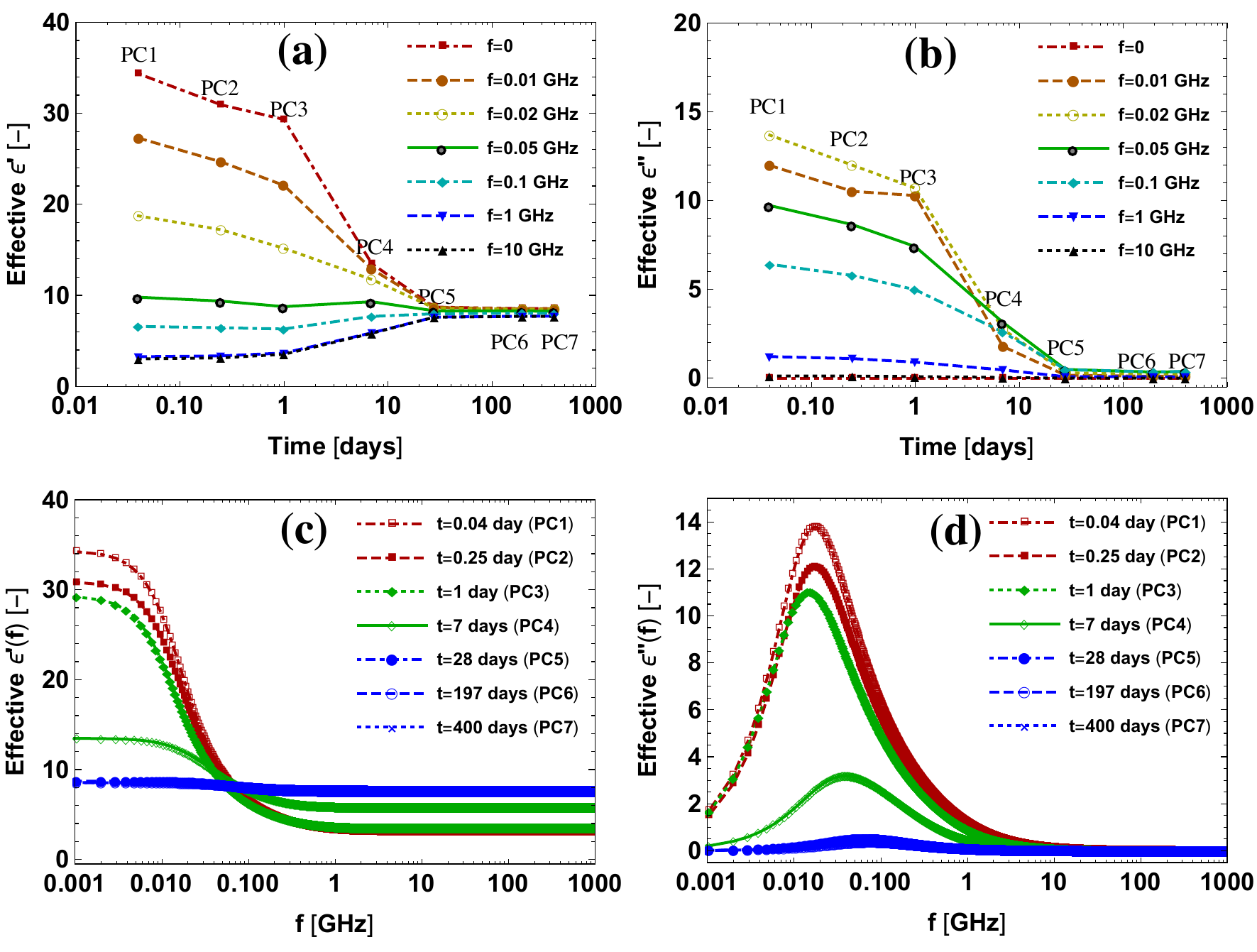}
 \caption{Effective permittivity $\epsilon = \Delta \chi+1=  \epsilon'- i\epsilon''$ of the cement paste with $w/c$=0.4. Time evolution of the (a) real $\epsilon'$ and (b) imaginary $\epsilon''$ parts at constant frequency $f$.
 Frequency-dependence of the (c) real $\epsilon'$ and (d) imaginary $\epsilon''$ parts at constant time.}
  \label{fgr:F9_eff_diel_perm}
\end{figure*}

The real $\epsilon'$ and imaginary $\epsilon''$ parts of the effective permittivity of the cement paste with a water-to-cement ratio ($w/c$) of 0.4 are shown in Fig. \ref{fgr:F9_eff_diel_perm}.
The frequency-dependence of both real and imaginary parts are very well pronounced in the range 0 to 1 GHz, especially at the early-ages ($t<10$).
The static values of the real part decrease with age.
The imaginary part reaches its maximum at a frequency in the range 0.01 to 0.2 GHz and decreases with age.
These results indicate that experimental measurements of effective permittivity can be used to discriminate the age of the cement paste or, using cross-property relations, to follow property development at early-age.

\section{Conclusions}

The dielectric spectra and frequency-dependent conductivities of bulk aqueous solutions representative of the pore solution found in cement-based materials were computed for the first time using classical molecular dynamics simulations.
Our main conclusions are as follows:

\begin{itemize}

\item Classical molecular dynamics simulations provide a powerful tool to compute the dielectric response of pore solutions with complex composition. Pore solutions in cement-based materials are cement type, age, and temperature-dependent. The dielectric spectra and frequency-dependent conductivities can be computed for various scenarios, which provide important input to the interpretation of the dielectric response of porous materials with complex fluid phase composition. 

\item The age-dependent composition of the pore solutions is reflected in the dielectric spectra.
In terms of static  susceptibility values, the difference between the maximum (PC3) and minimum (PC6) is of approximately 20.
This result shows that accounting for the composition variability of the pore is crucial to obtain the precise dielectric response of cement-based materials. 
The results provided in this paper can readily be used to upscale the dielectric response of ordinary Portland cement-based materials. Further studies are, though, necessary to quantify the dielectric response of highly confined water in the micropores found in C-S-H and AF-phases encountered in these materials. Again, molecular dynamics simulations can be a powerful tool in such an analysis.

\item Cole-Cole fits provide a reasonable description of the dielectric spectra of the pore solution studied. For pore solutions at the early ages (PC1-PC4), the Debye model seems sufficient to capture the real and imaginary parts of the spectrum (ie. $\alpha$ parameter in Cole-Cole model close to 0). As expected, the water contribution to the dielectric spectra is the most significant. The ion-ion and ion-water contributions are one to two orders inferior to the water contribution.

\item The age-dependent dielectric response of the pore solution can be used to upscale the dielectric response of cement pastes using effective medium approaches.
Both the static values of the real part and the peak in the imaginary part when plotted against the frequency decreases with the age.
These results show that the measurement of the (effective) dielectric response can be deployed to discriminate the age of cement pastes.

\end{itemize}

Our results can be used in the interpretation of dielectric probing of cement-based materials, enhancing the performance of non-destructive monitoring of concrete structures conditions such as radar testing and improving the confidence in the corresponding results.
Improvements on the detection of cracking and the diagnostic of structures pathologies (e.g. alkali-silica reactions, delayed ettringite formation, etc) can be envisioned using HF-EM techniques.
The next step will be focused on the modeling of the dielectric profile \cite{bonthuis_dielectric_2011} of confined electrolytes in cement hydrates micro- and meso-pores. 
This will bring fundamental knowledge about the electromagnetic properties of confined (bound) water in cement-based materials micro- and meso-pores, which is still not understood. Thus, a more realistic model of HF-EM properties of cement-based material could then be developed.

\bibliography{mybib0}

\end{document}


\preprint{APS/123-QED}

\title[Supporting Information]{Details on simulation procedures and results for computation of correlation functions}

\author{Tulio Honorio}
\affiliation{LMT, ENS Paris-Saclay, CNRS, Universit\'{e} Paris-Saclay, 94235 Cachan, France}%

\author{Thierry Bore}%
\affiliation{School of Civil Engineering, The University of Queensland, Brisbane, Australia}

\author{Farid Benboudjema}
\affiliation{LMT, ENS Paris-Saclay, CNRS, Universit\'{e} Paris-Saclay, Cachan, France}

\author{Eric Vourc'h}
\author{Mehdi Ferhat}
\affiliation{ 
SATIE, UMR CNRS 8029, ENS Paris Saclay, Universit\'{e} Paris Saclay, Cachan, France}

\date{\today}

\maketitle


\section{Force fields for molecular model of pore solutions}

We recall the force fields parameters described in a previous work \cite{honorio_pore_2019}.
The interactions among all the species in the system are modeled by a sum of non-bonded (van der Waals and electrostatic) and bonded (angle and bonds for OH groups) interactions.  
The van der Waals interactions are described either by the Lennard-Jones (12-6) potential (for interactions between water, hydroxide and monovalent ions):

\begin{equation}
 U^{VdW}_{LJ}(r_{ij})= \sum_{i \neq j} 4 \epsilon_{LJ} \left[  \left( \frac{\sigma_{LJ}}{r_{ij} } \right)^{12} - \left( \frac{\sigma_{LJ}}{r_{ij} }\right)^6 \right]  \label{eq:LJ}
\end{equation}

\noindent or by the Buckingham potential (for interactions involving sulfates): 

\begin{equation}
 U^{VdW}_{B}(r_{ij})= \sum_{i \neq j}  \left[A_B e^{-r_{ij} /\rho_B} - \frac{C_B}{r_{ij} ^6}  \right]  \label{eq:Buck}
\end{equation}

\noindent where $r_{ij}$ is the distance between the particles $i$ and $j$;  $\epsilon_{LJ}$, $\sigma_{LJ}$, $A_B$, $\rho_B$ and $C_B$ are empirical parameters. 
The parameters used for the Buckingham and the Lennard-Jones potentials are gathered in Tabs. \ref{tab:FF_species_LJ} and \ref{tab:FF_species_B} \cite{honorio_pore_2019}.
The table latter shows that for the interactions described by Buckingham potentials, all pair interactions are explicitly defined.
For the interactions described by Lennard-Jones interactions, Lorentz-Berthelot mixing rule is used for two dissimilar non-bonded atoms. 

The electrostatic contribution is described by the Coulomb potential:

\begin{equation}
U^{Coul}(r_{ij} )= \frac{e^2}{4 \pi  \epsilon_0 } \sum_{i \neq j} \frac{q_i q_j}{r_{ij}}  \label{eq:Coul}
\end{equation}

\noindent where $q_i$ is the partial charge of a particle $i$, $e$ is the elementary charge, and  $\epsilon_0=8.85419 \times 10^{-12}$ F/m is the dielectric permittivity of the free space. The partial charges deployed are gathered in Tab. \ref{tab:FF_partial_charges}.

Water and hydroxide bonds as well as water angles are constrained by SHAKE algorithm. The the equilibrium distance and angles, in this case, are $r_0$=~1~\r{A} and $\theta_0$~=~109.47~$^{\circ}$, respectively.
Sulfate ions are constrained using RIGID algorithm in LAMMPS.

\begin{table}
\small
\caption{\label{tab:FF_species_LJ} Lennard-Jones parameters for interactions between water, hydroxide and ions.}
\begin{ruledtabular}
\begin{tabular}{c|c|c|c}
 Pairs  &  $\epsilon_{LJ}$ [kJ.mol$^{-1}$] & $\sigma_{LJ}$ [\r{A}] & ref.  \\ 
\hline
H$_w$-H$_w$ &   -  & - & \footnotemark[1]\\   
O$_w$-O$_w$ &  0.650 & 3.166 &  \footnotemark[1] \\  
\hline
Na$^+$-Na$^+$ & 0.5443 & 2.350 & \footnotemark[2] \\
K$^+$-K$^+$   & 0.4184 & 3.742 &  \footnotemark[3]\\
Ca$^{2+}$-Ca$^{2+}$	& 0.4184 & 3.224 & \footnotemark[3]  \\
\hline
O$_H$-O$_H$ & 0.650 & 3.166  & \footnotemark[4] \\
H$_H$-H$_H$ &  &   & \footnotemark[4] \\
\end{tabular}
\end{ruledtabular}
\footnotetext[1]{Berendsen et al. \cite{berendsen_missing_1987}.}
\footnotetext[2]{Smith et al. \cite{smith_computer_1994}.}
\footnotetext[3]{Koneshan et al. \cite{koneshan_solvent_1998}.}
\footnotetext[4]{Brodskaya et al. \cite{brodskaya_computer_2003}.}
\end{table}

\begin{table}
\small
\caption{\label{tab:FF_species_B} Buckingham potential parameters for interactions involving sulfates.}
\begin{ruledtabular}
\begin{tabular}{c|c|c|c|c}
 Pairs &  $A_{B}$ [kJ.mol$^{-1}$] & $\rho_{B}$ [\r{A}] & C$_B$ [kJ.mol$^{-1}$.\r{A}$^6$] & ref. \\ 
\hline
O$_s$-O$_s$  & 570918.94 	& 0.2000 & 0	& \footnotemark[1]\\  
O$_s$-O$_w$  & 69636.03 	& 0.2649 & 0	&  \footnotemark[1]\\  
O$_s$-Na$^+$ & 3908.49 	& 0.2955 & 0	&  \footnotemark[1] \\  
O$_s$-K$^+$  & 8600.68 	& 0.2971 & 0		&  \footnotemark[1]\\  
O$_s$-Ca$^{2+}$ & 12928.54 	& 0.283474 & 0	&  \footnotemark[2] \\
O$_s$-O$_H$ & 69636.03 	& 0.2649 & 0	&  \footnotemark[1] \\   
S-S 	&  - 	& - & -	&  \footnotemark[1]\\
\end{tabular}
\end{ruledtabular}
\footnotetext[1]{Allan et al. \cite{allan_calculated_1993}.}
\footnotetext[2]{Byrne et al.  \cite{byrne_computational_2017}.}
\end{table}

\begin{table}
\small
\caption{\label{tab:FF_partial_charges} Partial charges.}
\begin{ruledtabular}
\begin{tabular}{c|c|c}
   & Charge [e] & ref.  \\ 
\hline
H$_w$ &  0.4238 & \footnotemark[1]\\   
O$_w$ & -0.8476 &  \footnotemark[1]\\  
\hline
Na$^+$ & 1.0 & \footnotemark[2] \\
K$^+$ & 1.0 &  \footnotemark[3]  \\
Ca$^{2+}$ & 2.0 &  \footnotemark[3]   \\
\hline
O$_H$ & -1.4238 & \footnotemark[4]  \\
H$_H$ & 0.4238 & \footnotemark[4]  \\
\hline 
O$_s$ & -0.84 & \footnotemark[5] \\ 
S	&  1.36 & \footnotemark[5]  \\  
\end{tabular}
\end{ruledtabular}
\footnotetext[1]{Berendsen et al. \cite{berendsen_missing_1987}.}
\footnotetext[2]{Smith et al. \cite{smith_computer_1994}.}
\footnotetext[3]{Koneshan et al. \cite{koneshan_solvent_1998}.}
\footnotetext[4]{Brodskaya et al. \cite{brodskaya_computer_2003}.}
\footnotetext[5]{Allan et al. \cite{allan_calculated_1993}.}
\end{table}

\section{Radial distribution functions}

\begin{figure*}
 \includegraphics[width=15cm]{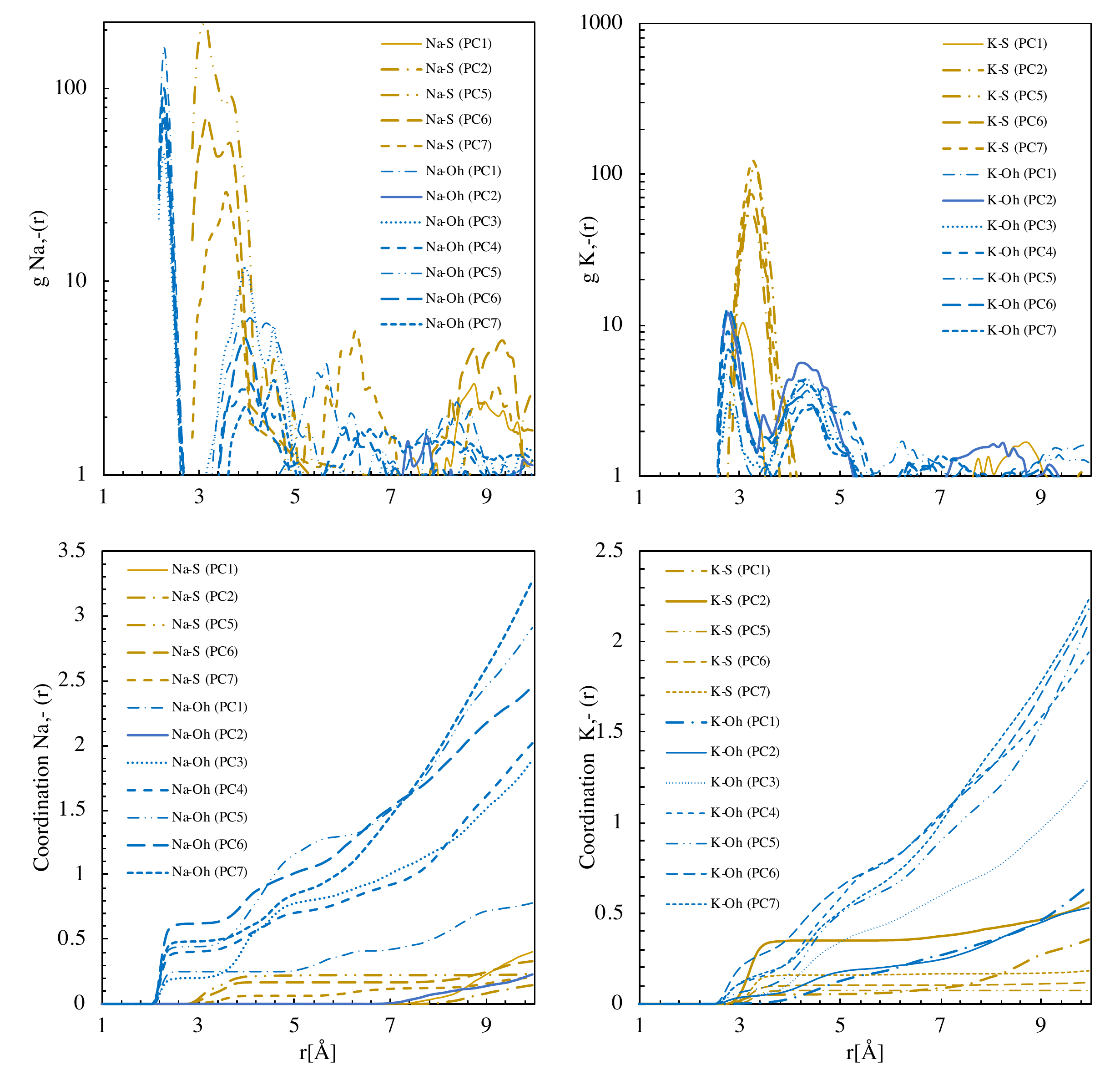}
 \caption{Radial distribution functions and cumulative numbers for K and Na ions. }
  \label{fgr:FSI1_RDF_Ions}
\end{figure*}

\begin{figure}
 \includegraphics[width=7cm]{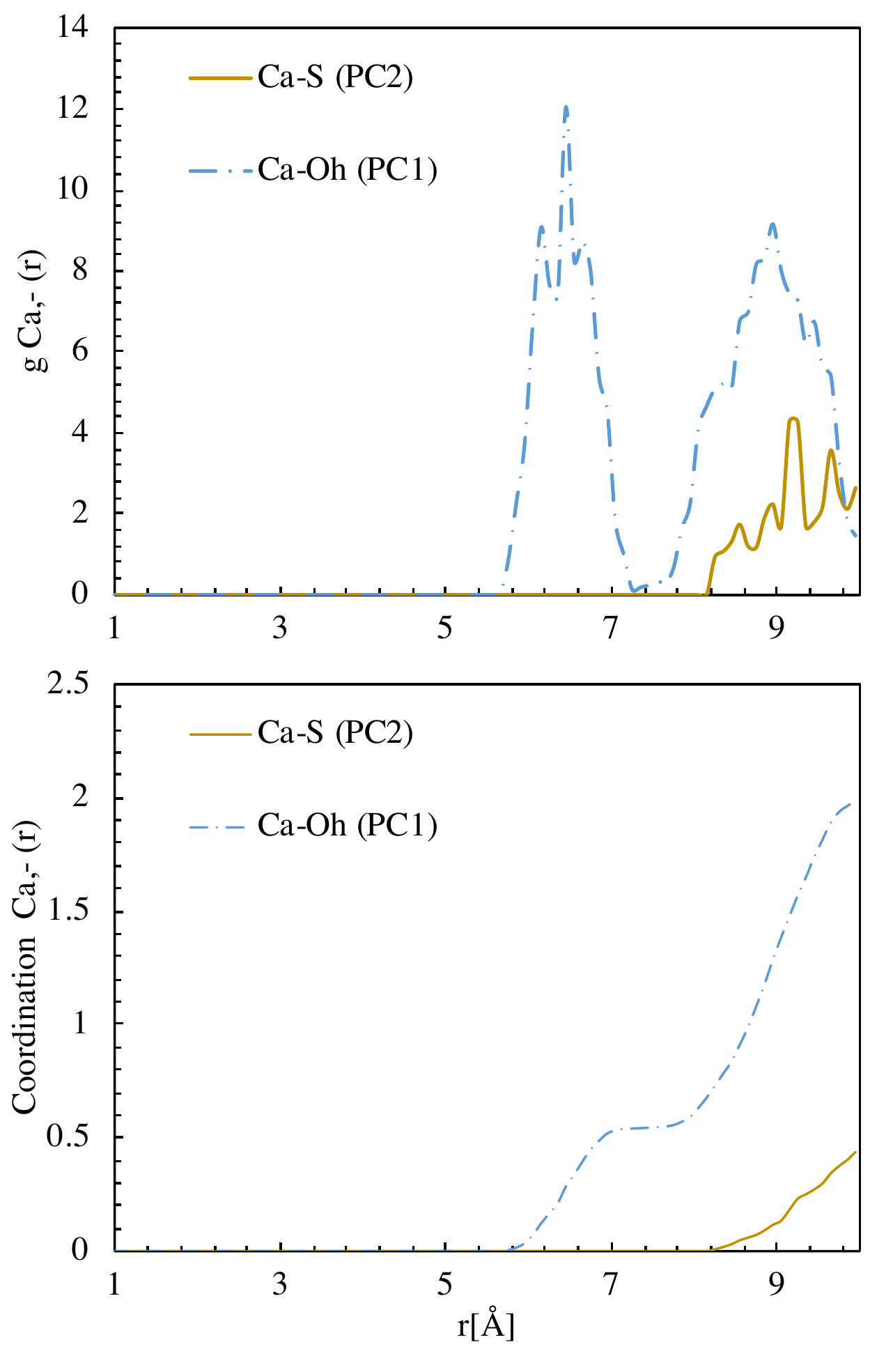}
 \caption{Radial distribution functions and cumulative numbers for Ca ions. }
  \label{fgr:FSI2_RDF_Ions_Ca}
\end{figure}

The ion-ion radial distribution functions used to compute these histograms of ion pair states are show in Fig.    \ref{fgr:FSI1_RDF_Ions} and  \ref{fgr:FSI2_RDF_Ions_Ca}.

\section{Correlation functions and integration}

\subsection{Polarization and ionic current flux vectors}
\begin{figure*}
 \includegraphics[width=15cm]{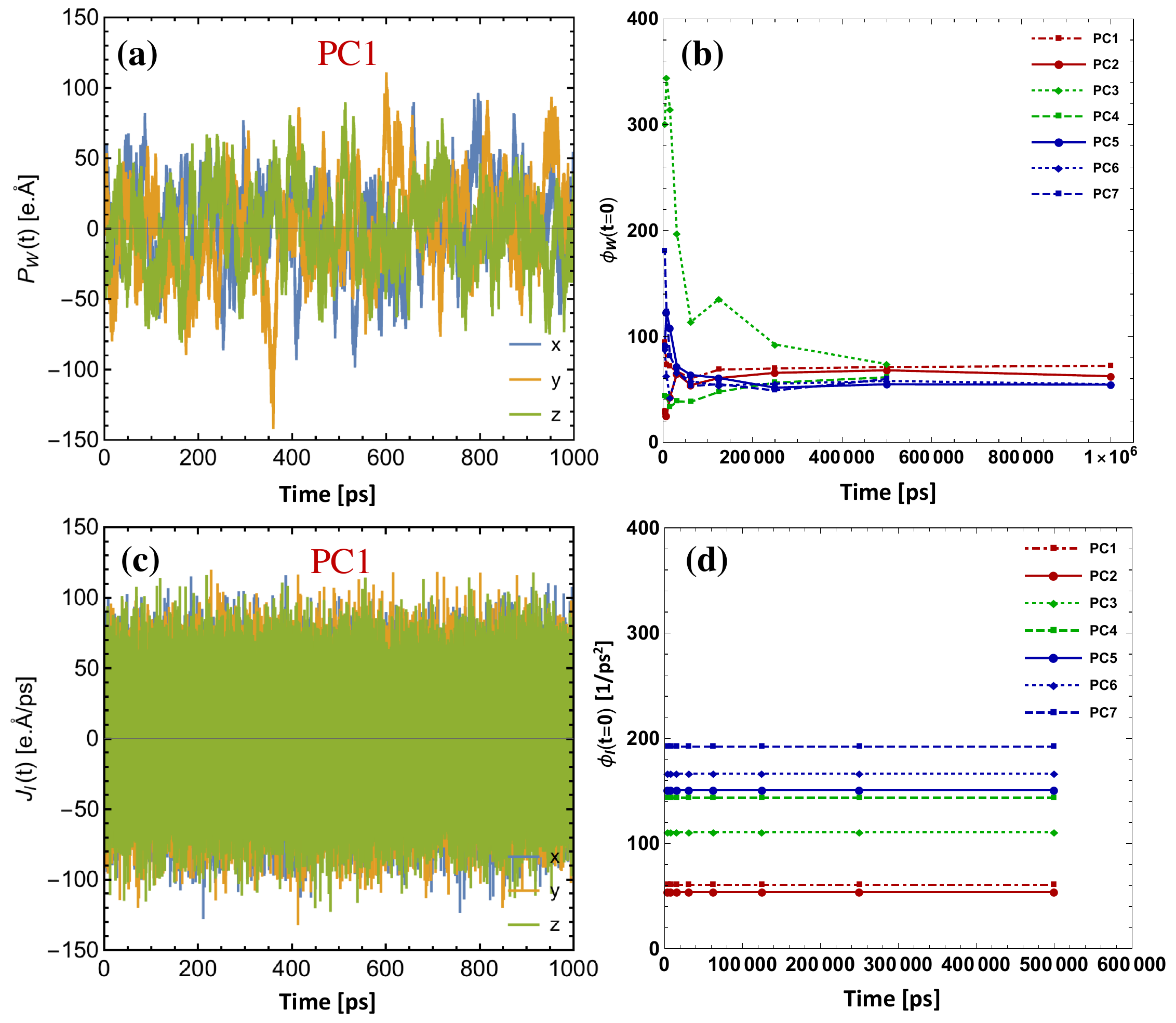}
 \caption{Components of (a) water polarization and (c) ionic current flux vectors for PC1 solution.
 Convergence of the (b) water auto-correlation function $\phi_W$ and (d) ion auto-correlation function $\phi_I$  both evaluated at $t=$0  as a function of the simulated time.}
  \label{fgr:FSI3_P_J_PC1}
\end{figure*}

The three components of the water polarization and ionic current flux vectors for PC1 solution are shown in 
Fig.  \ref{fgr:FSI3_P_J_PC1} (a) and (c).
The values of the auto-correlation functions at $t=$0 are used a a proxy to test the convergence. 
For all solution sutdied, the water auto-correlation function $\phi_W(0)$ converges for a simulation time of at least 0.5~ns (Fig.  \ref{fgr:FSI3_P_J_PC1} (b)), whereas the ion auto-correlation function $\phi_I(0)$ converges in a few ps.
These results indicates that a 1~ns simulation is reasonably sufficient to to compute the correlations functions of the polarization and ionic current.

\subsection{Absorption spectrum estimates}

To validate the simulation results against experimental data, one may compute specific spectra according to the vibrations experienced by the water molecules in the electrolytes \cite{praprotnik_molecular_2005}.
For example, the libration motions caused by the H-bond network in liquid bulk water leads to a broad band centered on approximately  685 cm$^{-1}$ in far infrared, Raman and  inelastic neutron scattering spectrum \cite{guillot_molecular_1991}.

The absorption coefficient per unit path length in infrared spectroscopy can be computed from the Fourier transform of the polarization autocorrelation function \cite{guillot_molecular_1991}:

\begin{equation}
\alpha(\omega)=\frac{4 \pi \omega \tanh \left[ \hbar \omega/(2 k_B T) \right]}{3 \hbar  n(\omega) c V}  \int^{\infty}_0 \left\langle \vec{P}(0). \vec{P}(t) \right\rangle e^{- i \omega t} dt \label{eq:I_IR}
\end{equation}

\noindent where $n(\omega)$ is the refractive index function of the angular pulsation $\omega=2 \pi f$, $\hbar$ is the reduced Planck constant and $c$ is the speed of light in vacuum.

\begin{figure}
 \includegraphics[width=8cm]{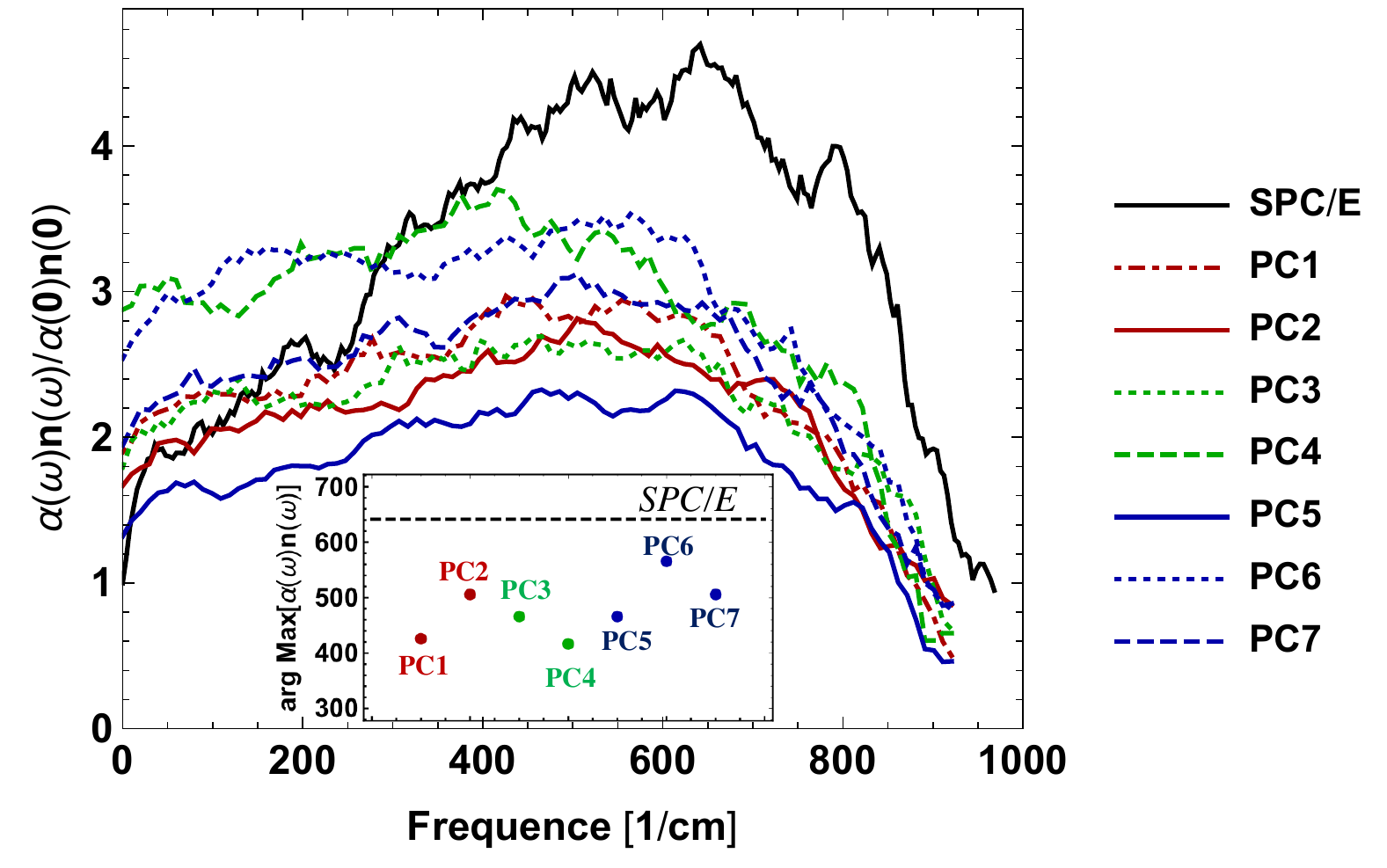}
 \caption{IR spectra of the pore solutions PC1-PC7 and SPC/E water. The inset shows the frequency to which each the spectrum is at its maximum. }
  \label{fgr:FSI4_IR_Spec}
\end{figure}

The IR spectra  obtained for of the pore solutions PC1-PC7 and SPC/E water are presented in Fig. \ref{fgr:FSI4_IR_Spec}. 
A moving average with window size of 25 fs is used to comput the spectra from the results obtained with Eq. \ref{eq:I_IR}.
No quantum corrections are considered.
The spectrum obtained for SPC/E water is consistent with the results of Guillot \cite{guillot_molecular_1991}.
The inset shows the frequency to which each the spectrum is at its maximum. 
For all pore solution, the maximum occurs at a frequency below the maximum frequency observed to SPC/E water.

\subsection{Integration of correlations functions }

The contributions of related to the water auto-correlation function $\phi_W(t)$ are the most relevant to the dielectric spectra of the solutions.
For SPC/E and the solution PC1-PC7, $\phi_W(t)$ first drops to zero between 25 and 50 ps.
The negative values are due to simulation noise \cite{rinne_dissecting_2014}.
The choice of the upper limit of integration in the range 25 and 50~ps lead to similar results. 
To compute the susceptibilities, we choose to integrate $\phi_W(t)$ up to 50~ps for all cases.  

The ion-water cross-correlation function $\phi_{IW}(t)$ exhibit a high oscillatory behavior, prone to be affected by simulation noise.
We adopt the strategy of Rinne et al. \cite{rinne_dissecting_2014} of integrating  $\phi_{IW}(t)$ up to the time threshold in which this function first drops below zero.
It must be noted that the resulting contribution to the susceptibility is is not very significant being two orders of magnitude below the contribution of the water auto-correlation. 
Therefore, the effects of the noise observed in $\phi_{IW}(t)$  are not relevant to the final dielectric response.

The ion auto-correlation function $\phi_{I}(t)$ exhibit a high oscillatory behavior that is strongly damped after few picoseconds.
In this study, $\phi_I(t)$ is integrated up to 1~ps when, for all solution studied, the amplitude of the oscillation are less than 10~ps$^{-2}$.

\subsection{Debye fits of the water contribution on the dielectric spectra}

\begin{figure*}
 \includegraphics[width=15cm]{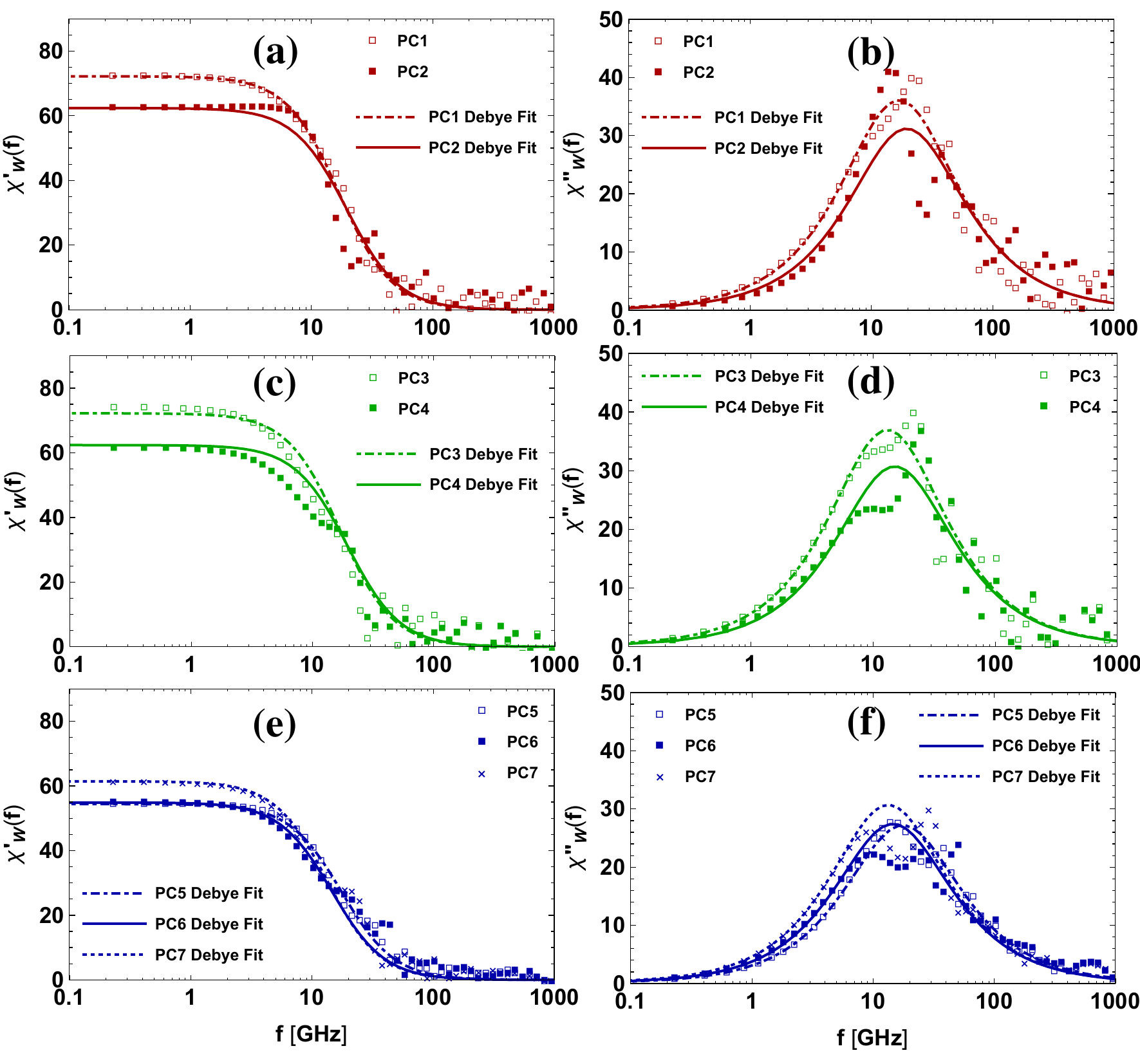}
 \caption{Dielectric spectra of the pore solutions PC1-PC7:  real (left) and imaginary (right) parts of the water contribution $\chi_W(f)$ to the dielectric susceptibility. (a)-(b) very early-age, (c)-(d) early-age, and (e)-(f) late ages pore solutions.}
  \label{fgr:FSI5_suscep_W}
\end{figure*}

The real and the imaginary parts of the water contribution $\chi_{W}(f)$ on the dielectric spectra of the pore solutions are shown in  Fig. \ref{fgr:FSI5_suscep_W}.
Simulation results were fitted with Debye model:

\begin{equation}
\Delta \chi(f)=\frac{\epsilon_{D}-\epsilon_{\infty}}{1+\left(i 2\pi f \tau_{D} \right)}+\epsilon_{\infty}-1  \label{eq:Debye}
\end{equation}

\noindent where $\epsilon_{D}$ and $\tau_{D}$ are, respectively, the Debye amplitude and characteristic relaxation time. We adopt $\epsilon_{\infty}$=1 since the force fields employed in MD simulation do not comprise atomic polarization.

Debye model fits quite well the water contribution $\chi_{W}(f)$ on dielectric spectra of all the solutions studied especially at low frequencies. Results at higher frequencies are prone to more noise.

\subsection{Finite size effects and NVE simulations}

Simulations were performed in PC1 systems with two- and four-fold the volume of the simulation box reported in the main text to quantify finite size effects.
The original simulation box with volume $V_0$ was duplicate then deformed so that a cubic box is obtained; the process is repeated to get a cubic box with volume $4V_0$.
The systems were then equilibrated during 0.5 ns in a NVT simulation followed by a 0.5 ns NVE simulation.
The production stage was performed in NVE ensemble, so that the effects of fictitious forces from the thermostat in NVT ensemble, which may affect the velocities and fluxes along MD trajectory, are excluded.

\begin{figure*}
 \includegraphics[width=15cm]{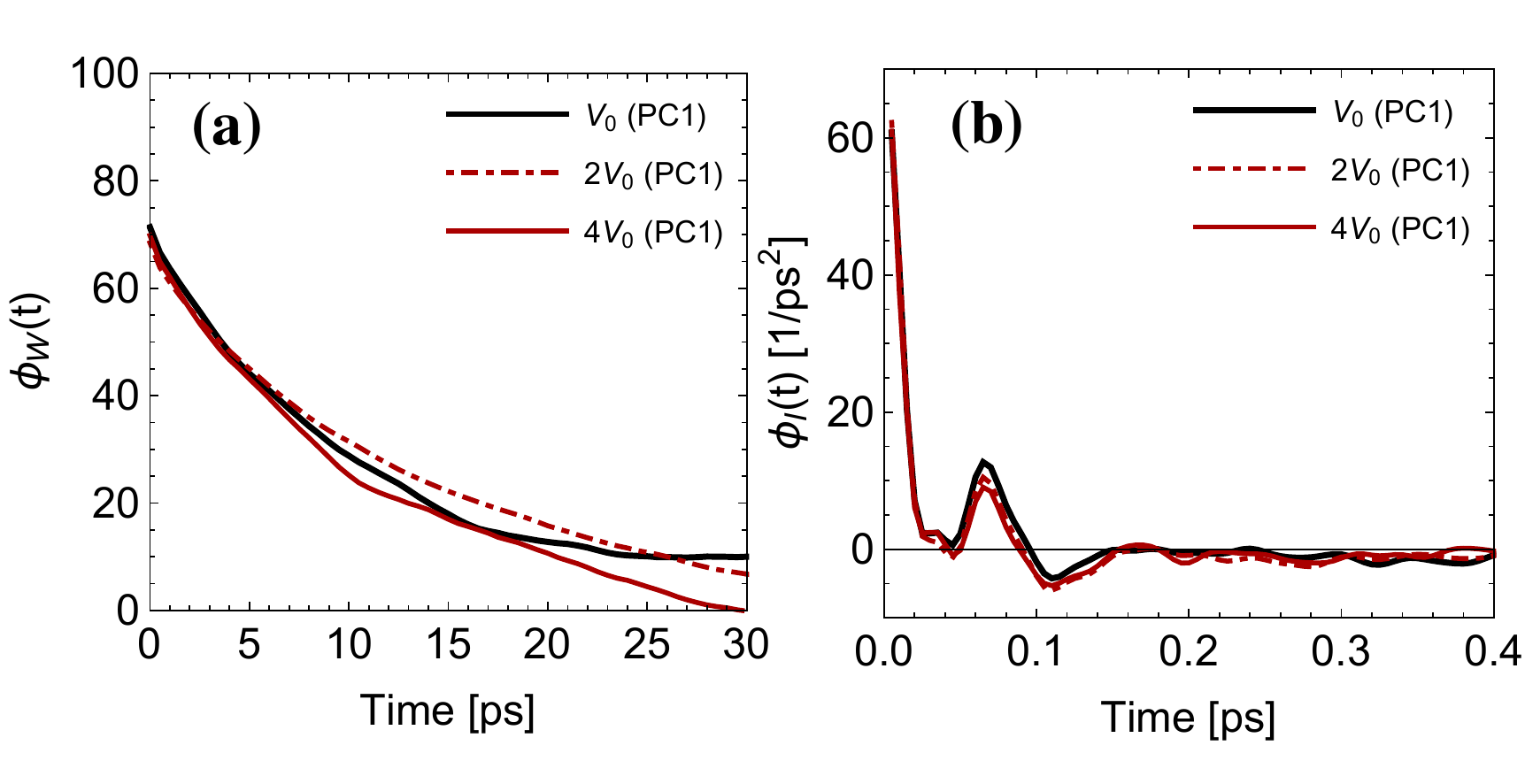}
 \caption{Auto-correlation function of the (a) water polarization and (b) ionic current for PC1 according to simulation box size. The original volume is denoted $V_0$.}
  \label{fgr:FSI6_phi_Siz}
  \end{figure*}

Figure~\ref{fgr:FSI6_phi_Siz} shows that the water and ion autocorrealtion function are not significantly affected by the thermostat and/or size effects.

\begin{figure*}
 \includegraphics[width=15cm]{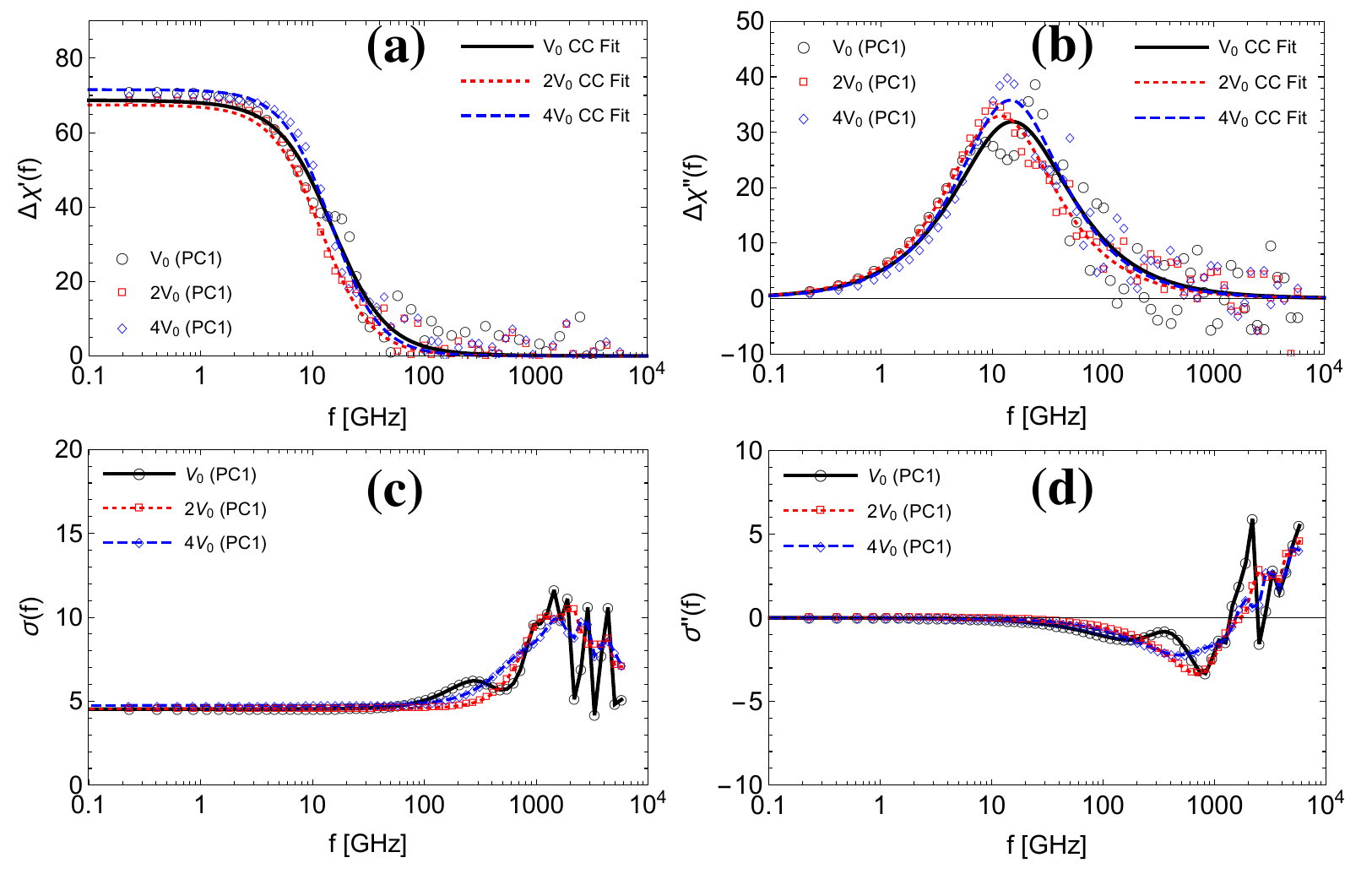}
 \caption{Susceptibility (top) and conductivity (bottom): real and imaginary contributions according to simulation box size. The original volume is denoted $V_0$. Cole-Cole fits are shown for the susceptibilities}
  \label{fgr:FSI7_sus_cond_Siz}
  \end{figure*}

Figure~\ref{fgr:FSI7_sus_cond_Siz} shows that both real and imaginary parts of the susceptibility and conductivity are not significantly affected by the thermostat and/or size effects up to frequencies of  1000 GHz.
To better sample higher frequencies, larger simulation boxes would be necessary.

\bibliography{mybib0}
